\documentclass[letterpaper,10pt]{IEEEtran}
\IEEEoverridecommandlockouts

\usepackage[colorlinks=true]{hyperref}
\hypersetup{urlcolor=blue,linkcolor=blue,citecolor=blue,colorlinks=true}
\usepackage{graphicx}	
\usepackage[space]{grffile}
\graphicspath{{./Figures/}}		
\usepackage{dcolumn}		
\usepackage{bm}				
\usepackage{float}		
\usepackage{amsmath, amsfonts}	
\usepackage{url}				
\usepackage{setspace}	
\usepackage{lineno}   		
\usepackage{siunitx}
\usepackage{color}
\newcommand*{\Scale}[2][4]{\scalebox{#1}{$#2$}}%

\title{\LARGE \bf
	Stochastic Models for Cochlear Instabilities
}

\author{Maurice Filo and Bassam Bamieh
	\thanks{This research was supported in part by NSF award CMMI-1363386.}
	\thanks{Maurice Filo and Bassam Bamieh are with the Department of Mechanical Engineering, University of California, Santa Barbara,
		Santa Barbara, California 93117, USA.       {\it\small filo@umail.ucsb.edu, bamieh@engineering.ucsb.edu}
}	}

\begin{document}
   \maketitle
   \thispagestyle{empty}
   \pagestyle{empty}
		
		\begin{abstract}
			We investigate instabilities in a stochastic mathematical model of cochlear dynamics. The cochlea is modeled as a spatio-temporal dynamical system made up of a  spatially distributed array of coupled oscillators, together with the cochlear amplification mechanism. We consider a setting where the cochlear amplifier has relatively small spatially and temporally varying stochastic parameters. It is shown that relatively small parameter variations (five to four orders of magnitude smaller than the nominal values) are sufficient to destabilize the dynamics and induce spontaneous oscillations. This extreme sensitivity of the cochlear dynamics  appears to be due to a combination of the local cochlear amplification mechanism, as well as the spatial coupling of the distributed resonators. We use an analysis technique which allows for a simulation-free prediction of the stability thresholds, as well as the statistics of the spontaneous oscillations. We verify our theoretical predictions using full non-linear stochastic simulations which appear to be in very good agreement with our theory.
		\end{abstract}


\maketitle



\section{\label{Sec: Introduction} Introduction}

The cochlea is a highly sensitive device that is capable of sensing sound waves across a broad spectrum of frequencies ($20-20000 \si{Hz}$) and across a wide range of sound intensities ranging from $0 \si{dB}$ (threshold of hearing) up to $120 \si{dB}$ (sound of a jet engine). The cochlea was believed to be a passive device that acts like a Fourier analyzer: each frequency causes a vibration at a particular location on the basilar membrane (BM). This mechanism was discovered by the Nobel Prize winner George von B\'ek\'esy who carried out his experiments on cochleae of human cadavers. However, in 1948, Thomas Gold hypothesized that the ear is rather an active device that has a component termed the cochlear amplifier. Although Gold's hypothesis was rejected by von B\'ek\'esy, David Kemp validated it thirty years later by measuring emissions from the ear. These emissions, termed otoacoustic emissions (OAEs) are sound waves that are produced by the cochlea and can be measured in the ear canal. 

It is widely accepted that the outer hair cells, anchored on the cochlear partition, are responsible for the active gain in the cochlea that produces these emissions. However, the underlying mechanism is still not well understood. For example, spontaneous otoacoustic emissions (SOAEs) -- emissions generated in the absence of any stimulus -- are studied in \cite{fruth2014active} and \cite{ku2008statistics}. The remarkable high sensitivity of the cochlea makes it vulnerable to stochastic perturbations that are believed to be the cause of these emissions. Particularly, in \cite{ku2008statistics}, the authors studied the instabilities that arise in a linear biomechanical cochlear model with spatially random active gain profiles that are static in time. In \cite{fruth2014active}, similar analysis was carried out on simplified cochlear models comprised of coupled active nonlinear oscillators. The randomness, or disorder, was introduced via static variations of a bifurcation parameter. In these previous works, the analysis was carried out through Monte Carlo simulations by studying the stability of different randomly generated active gain (or bifurcation) profiles. 

In this paper, we carry out a \textit{simulation-free} stability analysis of the linearized dynamics of a nonlinear model of the cochlea. Our analysis employs structured stochastic uncertainty theory (\cite{bamieh2018structured}, \cite{filo2018structured}, \cite{lu2002mean}, \cite{elia2005remote}) rather than Monte Carlo simulations, where the active gain is stochastic in space and time and may have a spatially-varying expectation and/or covariance. It turns out that letting the active gain be a stochastic process puts the model in a standard setting of linear time-invariant (LTI) systems in feedback with a diagonal stochastic process that enters the dynamics multiplicatively (see Figure~\ref{Fig: Feedback Block Diagram}). This analysis allows us to predict the locations on the BM where the dynamics are more likely to destabilize due to the underlying uncertainties. It also provides a bound on the variance of the perturbations allowed such that stability is maintained. 

The rest of the paper is organized as follows: we start by providing a brief description of a class of biomechanical models of the cochlea in section \ref{Section: Brief Model Description}. Then, in section \ref{Section: DSS Formulation}, we recast this class of models in a descriptor state space (DSS) form using operator language (i.e. in continuous space-time). In section \ref{Section: Active Gain Uncertainties}, we reformulate the DSS form in a standard setting that is particularly useful to carry out our stochastic uncertainty analysis. We also provide the conditions for mean-square stability (MSS). In section \ref{Section: Cochlear Instabilities}, we present the numerical results of the possible instabilities caused by stochastic gain profiles with different statistic properties. To validate our analysis, we show a stochastic simulation for the full nonlinear model in section~\ref{Section: Simulations}. Finally, before we conclude, we give a discussion in section~\ref{Section: Discussion} to give a physical interpretation of our results and provide some comments on previous works.

\section{Biomechanical Model of the Cochlea}
Throughout the literature, cochlear modeling attempts varied depending on two main factors. The first is concerned with the degree of biological realism of the mathematical model. This is realized by the incorporation of various biological structures (\cite{geisler1995cochlear}, \cite{lamar2006signal}, \cite{neely1986model}) and the dimensionality of the fluid filling the cochlear chambers (\cite{steele1979comparison}, \cite{givelberg2003comprehensive}). The second factor is concerned with the computational aspect of the models. Different numerical methods were devised to approach the spatio-temporal nature of the cochlea (\cite{neely1981finite}, \cite{elliott2013wave}). Particularly, \cite{elliott2007state} used a finite difference method developed in \cite{neely1981finite} to discretize space and formulate the model in state space form. Moreover, computationally efficient methods and model reduction techniques were developed for fast simulations of cochlear response (\cite{bertaccini2011fast}, \cite{filo2016order}). This section starts by describing the mathematical model adopted in this paper. Then, we reformulate the latter in a continuous space-time descriptor state space form, using operator language. This form has two advantages: (a) it encompasses a wider class of cochlear models and (b) it makes the dynamics more transparent by treating the exact model and its finite dimensional approximation (i.e. discretizing space by some numerical method) separately \cite{filo2017topics}.

\subsection{Mathematical Model Description} \label{Section: Brief Model Description}
\begin{figure}[h]
	\centering
	\begin{tabular}{l}
		\includegraphics[scale = 0.7]{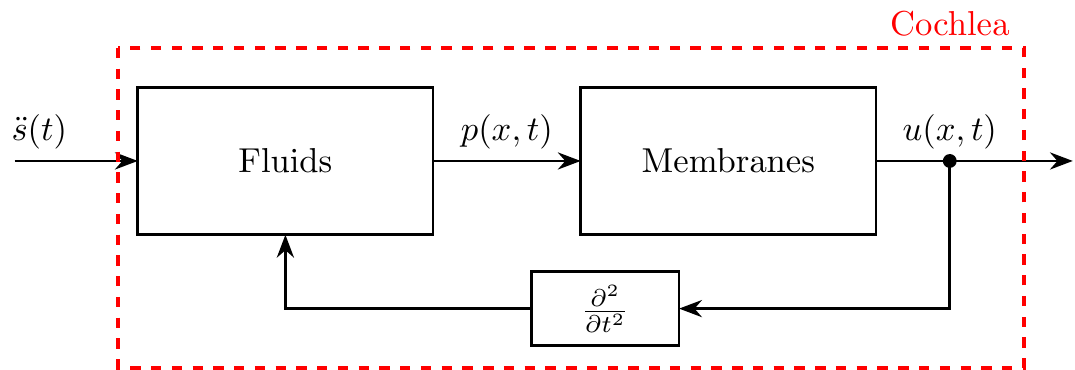} \\  \footnotesize{(a) Block Diagram of the Cochlea}  \\ 
		\includegraphics[scale = 0.7]{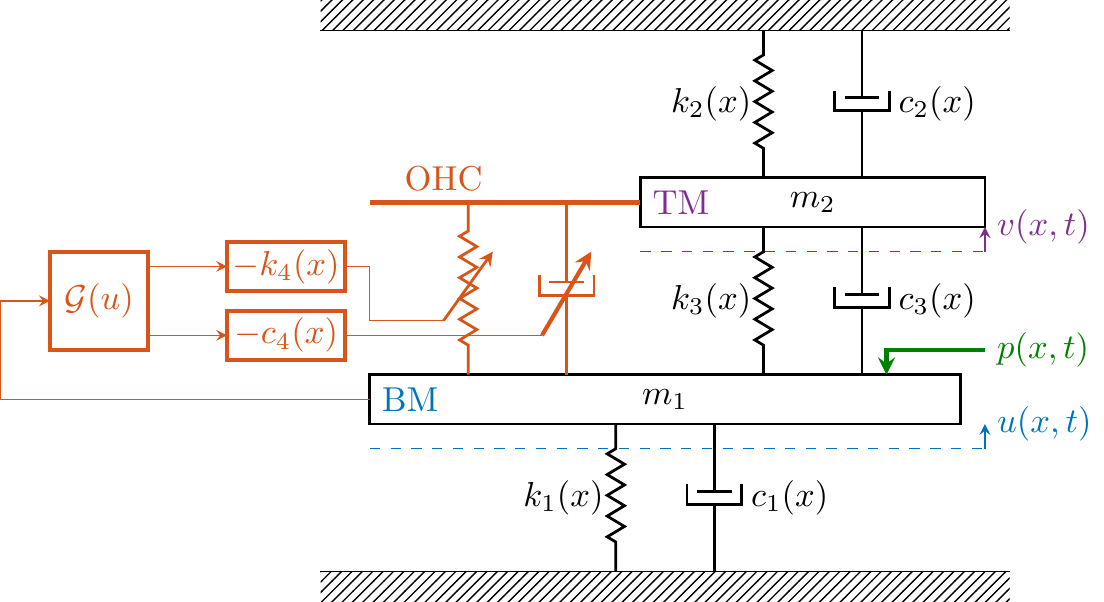} \\ \footnotesize{(b) Detailed Schematic Representing the Membranes Block }
	\end{tabular}
	\caption{\footnotesize{(a) The cochlea processes the acceleration of the stapes $\ddot s(t)$, in two stages, to produce the vibrations at every location of the BM, $u(x,t)$. The first stage is governed by the fluid that is stimulated by both the stapes and BM accelerations to yield a pressure $p(x,t)$ acting on every location of the BM. The second stage is governed by the dynamics of the membranes. The two stages are in feedback through the BM acceleration. (b) This figure is a schematic of a cross section (at a location $x$) of the cochlear partition showing the membranes governing the dynamics of the micro-mechanical stage. The spatially varying parameters $m_i$, $c_i(x)$ and $k_i(x)$ are the mass, damping coefficient and stiffness of the BM and TM for $i = 1$ and $2$, respectively. Furthermore, $c_3(x)$ and $k_3(x)$ are the mutual damping coefficient and stiffness, respectively; while $c_4(x)$ and $k_4(x)$ are the damping coefficient and stiffness associated with the active feedback gain from the outer hair cells (OHC) to the BM. The spring and damper between the BM and the OHC have variable negative values to capture the effect of the active force acting only on the BM without any direct effect on the TM. Their values depend on the the BM displacement $u$ via the nonlinear gain $\mathcal G(u)$. Equation(\ref{Eqn: Micromechanics}) describes the underlying dynamics.}}
	\label{Fig: Block Diagram of the Ear}
\end{figure}

The mathematical model can be divided into two main blocks as illustrated in Figure~\ref{Fig: Block Diagram of the Ear}(a). For a detailed derivation of the governing mechanics, refer to \cite{elliott2007state} and \cite{filo2016order} for a one and two dimensional modeling of the fluid stage, respectively. 
The fluid block, commonly referred to as the macro-mechanical stage, is linear and memoryless under the appropriate assumptions and approximations (refer to Appendix-\ref{Section: Mass Operators}). This block introduces spatial coupling along the different locations on the BM. Its output is the pressure $p(x,t)$ acting on each location of the BM. The governing equation can be written as a general expression, regardless of the dimensionality of the fluid and the numerical method used, as
\begin{equation} \label{Eqn: BM Pressure}
	p(x,t) = -[\mathcal M_f \ddot u](x,t) - [\mathcal M_s \ddot s](t),
\end{equation} 
where $\ddot{  }$ represents the second time derivative operation, and $\mathcal M_f$ and $\mathcal M_s$ are linear spatial operators associated with the fluid and stapes mass, respectively. Refer to Appendix-\ref{Section: Mass Operators} for a more detailed discussion of these mass operators and their finite dimensional approximations as matrices $M_f$ and $M_s$, respectively. The second block, commonly referred to as the micro-mechanical stage, takes the distributed pressure $p(x,t)$ as an input to produce the BM vibrations $u(x,t)$ at every location according to the following differential equations
\begin{equation} \label{Eqn: Micromechanics}
	\begin{aligned}
		\begin{bmatrix} \frac{g}{b} m_1 & 0 \\ 0 & m_2 \end{bmatrix} 
		\begin{bmatrix} \ddot u \\ \ddot v \end{bmatrix} + 
		\begin{bmatrix} \frac{g}{b}(c_1 + c_3 - \mathcal G(u)c_4) & \mathcal G(u) c_4 - c_3 \\ -\frac{g}{b}c_3 & c_2 + c_3 \end{bmatrix}
		\begin{bmatrix} \dot u \\ \dot v \end{bmatrix} \\ +
		\begin{bmatrix} \frac{g}{b}(k_1 + k_3 - \mathcal G(u)k_4) & \mathcal G(u) k_4 - k_3 \\ -\frac{g}{b}k_3 & k_2 + k_3 \end{bmatrix}
		\begin{bmatrix} u \\ v \end{bmatrix} =
		\begin{bmatrix} p \\ 0 \end{bmatrix},
	\end{aligned}
\end{equation}
where $v(x,t)$ is the tectorial membrane (TM) vibration (refer to Figure~\ref{Fig: Block Diagram of the Ear}(b)). Note that the space and time variables $(x,t)$ are dropped where necessary for notational compactness.  The constant $b$ is the ratio of the average to maximum vibration along the width of the BM, and $g$ is the BM to outer hair cells lever gain. Refer to \cite{neely1986model} for a detailed explanation of the parameters. Finally, $\mathcal G$ is the nonlinear active gain operator that captures the active nature of the outer hair cells, commonly referred to as the cochlear amplifier. In the spirit of \cite{lamar2006signal}, the action of $\mathcal G$ on a distributed BM displacement profile $u$ is given by
\begin{equation} \label{Eqn: Nonlinear Gain}
	\left[\mathcal G(u)\right](x,t) = \frac{\gamma(x)}{1 + \theta \left[\Phi_\eta\left(\frac{u^2}{R^2}\right)\right](x,t)},
\end{equation}
where the gain coefficient $\gamma(x)$ represents the gain at a location $x$, in the absence of any stimulus ($u(x,t) = 0$). The constants $\theta$ and $R$ are the nonlinear coupling coefficient and BM displacement normalization factor, respectively. The operator $\Phi_{\eta}$ is a normalized Gaussian operator such that its action on $u$ is defined as
\begin{align} 
	[\Phi_\eta(u)](x,t) &:= \frac{\int_0^L \phi_{\eta}(x-\xi) u(\xi,t) d\xi}{\int_0^L \phi_{\eta}(x-\xi)d\xi}; \label{Eqn: Guassian Weighing Operator}\\
	\phi_{\eta}(x) &:= \frac{1}{\eta\sqrt{2\pi}}e^{\frac{-x^2}{2\eta^2}}\label{Eqn: Gaussian Kernel},
\end{align}
where $L$ is the length of the BM and $\phi_{\eta}$ is the Gaussian kernel with a width $\eta$.
Note that $\eta = 0.5345 \si{mm}$ corresponds to the equivalent rectangular bandwidth on the BM (refer to Appendix-\ref{Section: ERB} for a detailed explanation).
Observe that the spatial coupling in the micro-mechanical stage appears only in the nonlinear active gain (\ref{Eqn: Nonlinear Gain}).

\subsection{Deterministic Descriptor State Space Formulation of the Linearized Dynamics in Continuous Space-Time} \label{Section: DSS Formulation}
This section gives a Descriptor State Space (DSS) formulation of the cochlear model described in (\ref{Eqn: BM Pressure}) and (\ref{Eqn: Micromechanics}). The DSS form is given for the linearized dynamics around the only fixed point which is the origin. 

It can be shown (Appendix-\ref{Section: System Linearization}) that the linearized dynamics can be achieved by simply replacing the nonlinear active gain $[\mathcal G(u)](x,t)$ in (\ref{Eqn: Micromechanics}) by its gain coefficient $\gamma(x)$. First, define the state space variable $\psi(x,t)$ in continuous space-time as
\begin{equation} \label{Eqn: State Space Variable}
\psi(x,t) := \begin{bmatrix} u(x,t) & v(x,t) & \dot u(x,t) & \dot v(x,t) \end{bmatrix}^T.
\end{equation}
Then the DSS form of the linearized dynamics is 
\begin{equation} \label{Eqn: Descriptor State Space Operator Form}
	\begin{aligned}
		\mathcal E \frac{\partial}{\partial t} \psi(x,t) &= \mathcal A_{\gamma} \psi(x,t) + \mathcal B \ddot s(t) \\
		u(x,t) &= \mathcal C \psi(x,t),
	\end{aligned}
\end{equation}
where $\mathcal E$, $\mathcal A_{\gamma}$ and $\mathcal B$ are matrices of linear spatial operators defined as follows
\begin{align*}
		&\mathcal E := 
		\begin{bmatrix}
			\mathcal I 		& 0				& 0											& 0					\\
			0				& \mathcal I	& 0											& 0					\\
			0				& 0				& \frac{g}{b}m_1 \mathcal I  + \mathcal M_f	& 0					\\
			0				& 0				& 0											& m_2 \mathcal I 	\\
		\end{bmatrix}; \quad \mathcal B := 
		\begin{bmatrix} 0 \\ 0 \\ -\mathcal M_s \\ 0 \end{bmatrix}; \\ 
		&\mathcal A_{\gamma} := \mathcal A_0 + \mathcal B_0 \gamma \mathcal C_0; \qquad 
		\mathcal C := \begin{bmatrix} \mathcal I & 0 & 0 & 0 \end{bmatrix}; \\
		&\Scale[0.97]{\mathcal A_0 := \resizebox{0.92\hsize}{!}{$
		\begin{bmatrix}
			0						& 0				& \mathcal I				& 0\\
			0						& 0				& 0							& \mathcal I\\
			-\frac{g}{b}(k_1+k_3)	& k_3			& -\frac{g}{b}(c_1+c_3)		& c_3\\
			\frac{g}{b}k_3			& -(k_2+k_3)	& \frac{g}{b}c_3			&-(c_2+c_3)	
		\end{bmatrix}$};}  \\
		&\mathcal B_0 := \begin{bmatrix} 0 & 0 & \mathcal I & 0 \end{bmatrix}^T; \qquad 
		\mathcal C_0 := \begin{bmatrix} \frac{g}{b}k_4 & -k_4 & \frac{g}{b}c_4 & -c_4 \end{bmatrix};
\end{align*}
and $\mathcal I $ is the identity operator. The equations in (\ref{Eqn: Descriptor State Space Operator Form}) represent a deterministic evolution differential equation and an output equation that provides the distributed displacement of the BM $u(x,t)$. Other outputs can be selected, such as the TM displacement, by appropriately constructing the $\mathcal C$ operator. In the subsequent section, we slightly modify the dynamical equations to account for stochastic perturbations in the gain coefficient $\gamma(x)$.

\section{Stochastic Uncertainties in the Active Gain} \label{Section: Active Gain Uncertainties}
This section investigates the Mean Square Stability (MSS, which we will formally define in section \ref{Section: Stochastic Feedback Interconnection}) of the linearized cochlear dynamics when the gain coefficient is a \textit{spatio-temporal stochastic process}. The stochastic gain coefficient, now denoted by $\gamma(x,t)$ to account for spatio-temporal perturbations, enters the dynamics (\ref{Eqn: Descriptor State Space Operator Form}) multiplicatively. We first reformulate the dynamics as an LTI system in feedback with a diagonal stochastic gain which is a standard setting in robust control theory\cite[Section 10.3]{zhou1996robust}. Then we carry out our MSS analysis based on \cite{filo2018structured}. By tracking the evolution of the instantaneous spatial covariances, MSS analysis allows us to predict the locations on the BM that are more likely to become unstable due to the underlying stochastic uncertainty. We conclude this section by defining and analyzing a linear operator, whose spectral radius provides a condition for MSS. 

\subsection{Stochastic Feedback Interconnection} \label{Section: Stochastic Feedback Interconnection}
The purpose of this section is to separate the stochastic portion of the gain coefficient in a feedback interconnection. We assume that $\gamma(x,t)$ is a \textit{spatio-temporal stochastic process} that is white in time (but may be colored in space), and whose expectation and covariance are independent of time. More precisely, let $\bar \gamma(x)$ be the expectation of $\gamma(x,t)$ and $\tilde \gamma(x,t)$ be a temporally independent, zero mean stochastic perturbation, such that
\begin{equation} \label{Eqn: Stochastic Gain}
	\begin{aligned}
		\gamma(x,t) &= \bar \gamma(x) + \epsilon \tilde \gamma(x,t), \\
		\text{with } & \left\{
		\begin{aligned}
			&\mathbb E[\gamma(x,t)] = \bar \gamma(x) \\
			&\mathbb E[\tilde \gamma(x,t) \tilde \gamma(\xi,\tau)] = \mathbf \Gamma(x,\xi) \delta(t-\tau)
		\end{aligned}\right. ~~ \forall t\geq 0,
	\end{aligned}
\end{equation}
where $\mathbb E[.]$ denotes the expectation, $\epsilon$ is a perturbation parameter, $\delta(t)$ is the Dirac Delta function, and $\mathbf \Gamma(x,\xi)$ is a positive semi-definite covariance kernel.
Substituting (\ref{Eqn: Stochastic Gain}) in (\ref{Eqn: Descriptor State Space Operator Form}) yields
\begin{equation} \label{Eqn: Stochastic Feedback}
	\begin{aligned}
		\mathcal E \frac{\partial}{\partial t}\psi(x,t) &= (\mathcal A_{\bar \gamma}  + \epsilon \mathcal B_0 \tilde \gamma \mathcal C_0) \psi(x,t) + \mathcal B \ddot s(t)\\
		u(x,t) &= \mathcal C \psi(x,t).
	\end{aligned}
\end{equation}
The evolution equation in (\ref{Eqn: Stochastic Feedback}) is a Stochastic Partial Differential Equation (SPDE) that is given an It\=o interpretation in the time variable. For more details on It\=o calculus, refer to \cite{oksendal2003stochastic}.

\begin{figure}[h]
	\centering
	\includegraphics[scale = 0.9]{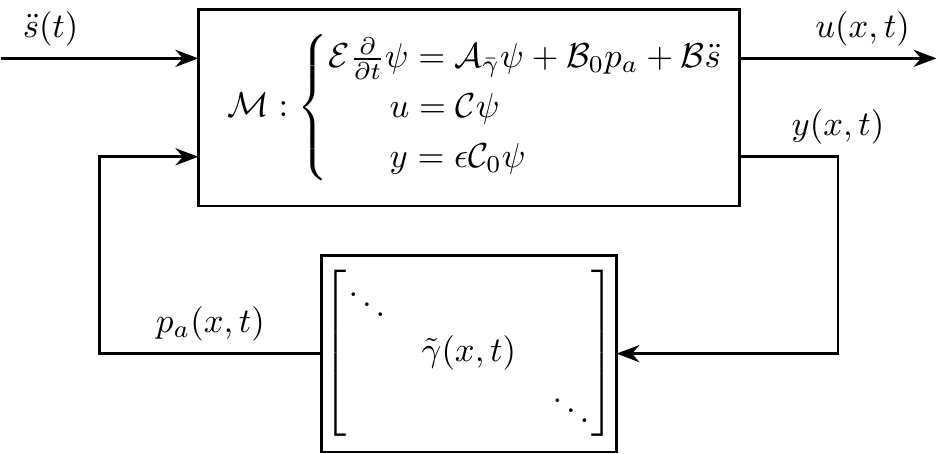}
	\caption{\footnotesize{The linearized cochlear model in feedback with multiplicative stochastic gain. The block to the top represents the deterministic portion of the linearized cochlear dynamics casted in a descriptor state space form. The feedback block is a diagonal spatial operator that represents the multiplicative stochastic gain. $y(x,t)$ is the differential vibration and velocity between the BM and TM as given by (\ref{Eqn: Secondary Output}). $p_a(x,t)$ is the active pressure that results from the stochastic component of the active gain.}}
	\label{Fig: Feedback Block Diagram}	
\end{figure}

Define a secondary output related to the difference in BM and TM displacements and velocities as
\begin{equation} \label{Eqn: Secondary Output}
	\begin{aligned}
		y(x,t) &:= \epsilon \mathcal C_0 \psi(x,t).
	\end{aligned}
\end{equation}
Furthermore, define the active feedback pressure resulting from the stochastic perturbations to be
\begin{equation} \label{Eqn: Feedback Pressure}
	p_a(x,t) := \tilde \gamma(x,t) y(x,t).
\end{equation}
Therefore, using (\ref{Eqn: Stochastic Feedback}), (\ref{Eqn: Secondary Output}) and (\ref{Eqn: Feedback Pressure}), construct the feedback block diagram depicted in Figure \ref{Fig: Feedback Block Diagram}.
This is a standard setting(\cite{filo2018structured} for structured stochastic uncertainty analysis, where the feedback gain is a diagonal spatial operator. This configuration is used to investigate the MSS of the cochlea which is formally defined next.

\textit{Definition}: The feedback system in Figure~\ref{Fig: Feedback Block Diagram} is MSS if, in the absence of an input (i.e. $\ddot s(t) = 0$), the state $\psi(x,t)$ and the active feedback pressure $p_a(x,t)$ have bounded variances for all time. 

Therefore, to study MSS, we need to track the temporal evolution of the variances and look at their steady state limits as $t$ goes to $+\infty$. This is the topic of the next subsection.

\subsection{Temporal Evolution of the Covariance Operators} \label{Section: Covariance Evolution}
This section tracks the time evolution of the covariance operators in the absence of any input (i.e. we set $\ddot s(t) = 0$ for the rest of the paper). We use the term covariance ``operators" rather than covariance matrices because the spatial variables $x$ and $\xi$ are continuous. After using some numerical method to discretize space, the covariance operators can be approximated by covariance matrices.   
With slight abuse of notation, we use the same symbol to denote the covariance operator and its associated kernel. Define the following instantaneous spatial covariance kernels
\begin{equation} \label{Eqn: Covariance Definitions}
	\begin{aligned}
		\mathcal X(x,\xi;t) &:= \mathbb E[\psi(x,t) \psi(\xi,t)] \\
		\mathcal Y(x,\xi;t) &:= \mathbb E[y(x,t) y(\xi,t)] \\
		\mathcal P(x,\xi;t) &:= \mathbb E[p_a(x,t) p_a(\xi,t)] \\
		\mathcal U(x,\xi;t) &:= \mathbb E[u(x,t) u(\xi,t)] \\
		\mathbf \Gamma(x,\xi) &:= \mathbb E[\tilde \gamma(x,t) \tilde \gamma(\xi,t)] \qquad \forall t\geq 0.
	\end{aligned}
\end{equation}
Given that the stochastic perturbations $\tilde \gamma$ are temporally independent, it can be shown \cite[Section V]{filo2018structured} that the time evolution of the covariance operators are governed by the following operator-valued, \textit{differential algebraic} equations
\begin{equation} \label{Eqn: Covariance Evolution Equations}
	\begin{aligned}
		\mathcal E\dot {\mathcal X} \mathcal E^* &= \mathcal A_{\bar \gamma} \mathcal X \mathcal E^* + \mathcal E \mathcal X \mathcal A_{\bar \gamma}^*  + \mathcal B_0 \mathcal P \mathcal B_0^* \\
		\mathcal Y &= \epsilon ^2 \mathcal C_0 \mathcal X \mathcal C_0^*\\
		\mathcal P &= \mathbf \Gamma \circ \mathcal Y,
	\end{aligned}
\end{equation}
where $*$ is the adjoint operation and $\circ$ is the Hadamard product; i.e. the element-by-element multiplication of the kernels $\mathcal P(x,\xi;t) = \mathbf \Gamma(x,\xi) \mathcal Y(x,\xi;t)$. 

In order to study the MSS, we need to look at the steady state limit of the covariances. We denote by the asymptotic limit of a covariance operator, when it exists, by an overbar. That is
\begin{equation} \label{Eqn: Covariance Limits}
	\begin{aligned}
		\bar {\mathcal X} := \lim_{t\to\infty} \mathcal X(t); ~~
		\bar {\mathcal Y} := \lim_{t\to\infty} \mathcal Y(t); ~~ 
		\bar {\mathcal P} := \lim_{t\to\infty} \mathcal P(t) .
	\end{aligned}
\end{equation}
At the steady state, the covariances become constant in time and thus their time derivatives go to zero. Hence, the steady state covariances, if they exist, are governed by the following operator-valued \textit{algebraic} equations:
\begin{equation} \label{Eqn: State State Covariance Equations}
	\begin{aligned}
		&\mathcal A_{\bar \gamma} \bar{\mathcal X} \mathcal E^* + \mathcal E \bar{\mathcal X} \mathcal A_{\bar \gamma}^*  + \mathcal B_0 \bar{\mathcal P} \mathcal B_0^* = 0 \\
		&\bar{\mathcal Y} = \epsilon^2 \mathcal C_0 \bar{\mathcal X} \mathcal C_0^*\\
		&\bar{\mathcal P} = \mathbf \Gamma \circ \bar{\mathcal Y}.
	\end{aligned}
\end{equation}
In the next section, we will use (\ref{Eqn: State State Covariance Equations}) to define a new operator as a tool to check the boundedness of the steady state covariances.

\subsection{Loop Gain Operator \& MSS}
Using (\ref{Eqn: State State Covariance Equations}), define the loop gain operator $\mathbb{L}_{\mathbf \Gamma}$, parametrized by the perturbation covariance $\mathbf \Gamma$, as 
\begin{equation} \label{Eqn: Loop Gain Operator}
	\begin{aligned}
		\mathbb{L}_{\mathbf \Gamma}(\bar{\mathcal P}_{\text{in}}) = \bar{\mathcal P}_{\text{out}}  \Longleftrightarrow  \left\{
		\begin{aligned}
			&\bar{\mathcal P}_{\text{out}} = \mathbf \Gamma \circ (\mathcal C_0 \bar{\mathcal X} \mathcal C_0^*) \\
			&\mathcal A_{\bar \gamma} \bar{\mathcal X} \mathcal E^* + \mathcal E \bar{\mathcal X} \mathcal A_{\bar \gamma}^*  + \mathcal B_0 \bar{\mathcal P}_{\text{in}} \mathcal B_0^* = 0.  
		\end{aligned} \right.
	\end{aligned}
\end{equation}
The MSS condition is given in terms of the spectral radius of the loop gain operator as explained next.

\textit{\textbf{Theorem}}: \textit{Consider the system in Figure~\ref{Fig: Feedback Block Diagram} where $\tilde \gamma$ is a temporally independent multiplicative noise, interpreted in the sense of It\=o, with instantaneous spatial covariance $\mathbf \Gamma$, and $\mathcal M$ is a stable causal LTI system. The feedback system is MSS if and only if the spectral radius of the loop gain operator is strictly less than one, i.e. 
\begin{equation} \label{Eqn: MSS Condition}
	\epsilon^2 \rho(\mathbb L_{\mathbf \Gamma}) < 1,
\end{equation}
where $\mathbb L_{\mathbf \Gamma}$ is defined in (\ref{Eqn: Loop Gain Operator}) and $\rho(\mathbb L_{\mathbf \Gamma})$ is its spectral radius.} 

The proof of this theorem is given in \cite{filo2018structured}. This theorem will be used to find an upper bound on the perturbation constant $\epsilon$ above which MSS is violated.

\subsection{Worst-Case Covariances}
The loop gain operator maps a covariance operator $\bar{\mathcal P}_{\text{in}}$ into another covariance operator $\bar{\mathcal P}_{\text{out}}$. Hence, the eigenvectors of $\mathbb L_{\mathbf \Gamma}$ are themselves operators. When a finite dimensional approximation of $\mathbb L_{\mathbf \Gamma}$ is carried out using some numerical method, these eigenvectors can be approximated as matrices.  We are particularly interested in the eigenvector (or eigen-operator) of $\mathbb L_{\mathbf \Gamma}$ associated with the largest eigenvalue because it has a significant meaning explained in this subsection. 

First, since the loop gain operator is a monotone operator \cite{bamieh2018structured}, it is guaranteed to have a real largest eigenvalue equal to $\rho(\mathbb L_{\mathbf \Gamma})$. It is also guaranteed that the eigen-operator associated with the largest eigenvalue is positive semidefinite, i.e. there exists a positive semidefinite covariance operator $\textbf{P}$ such that
\begin{equation} \label{Eqn: Worst Case Covariance}
	\mathbb L_{\mathbf \Gamma}(\textbf{P}) = \rho(\mathbb L_{\mathbf \Gamma}) \textbf{P}.
\end{equation}
Note that $\textbf{P}$ is the operator counterpart of the Perron-Frobenius eigenvector for matrices with non-negative entries. Refer to \cite[Thm 2.3]{bamieh2018structured} for a proof of the aforementioned guarantees. If the stability condition (\ref{Eqn: MSS Condition}) is violated, $\textbf{P}$ will be the covariance mode that has the highest growth rate, hence the name ``worst-case" covariance. This provides information about the locations on the BM that are more likely to destabilize due to the stochastic perturbations of the gain. Particularly, since we are interested in the instabilities at the BM, the worst-case covariance of the BM vibrations, denoted by $\mathbf U$, can be computed by propagating the worst-case pressure covariance $\textbf{P}$ through the cochlear dynamics (at steady state) as follows
\begin{equation} \label{Eqn: BM Worst Case Covariance}
	\begin{aligned}
		&\mathcal A_{\bar \gamma} \textbf{X} \mathcal E^* + \mathcal E \textbf{X} \mathcal A_{\bar \gamma}^*  + \mathcal B_0 \textbf{P} \mathcal B_0^* = 0  \\
		&\textbf{U} = \mathcal C \textbf{X} \mathcal C^*,
	\end{aligned}
\end{equation}
where $\textbf{X}$ denotes the worst-case covariance operator corresponding to the state space variable $\psi$.

\section{Instabilities in Linearized Cochlear Dynamics} \label{Section: Cochlear Instabilities}
This section contains the main results on the effects of stochastic uncertainties on cochlear instabilities. The analysis is carried out for three different scenarios of the perturbation covariance $\mathbf \Gamma(x,\xi)$:
\begin{itemize}
	\item $\textbf{S}_1$: spatially uncorrelated uncertainties, i.e. ${\mathbf \Gamma(x,\xi) = \delta(x-\xi)}$
	\item $\textbf{S}_2$: spatially correlated uncertainties with a correlation length $\lambda$, i.e. $\mathbf \Gamma(x,\xi) = \phi_{\lambda}(x-\xi)$
	\item $\textbf{S}_3$: spatially localized and uncorrelated uncertainties, i.e. $\mathbf \Gamma(x,\xi) = \phi_{\sigma}(x-\mu) \delta(x-\xi)$,
\end{itemize}
where $\phi_\lambda$ and $\phi_{\sigma}$ are the Gaussian kernels defined in (\ref{Eqn: Gaussian Kernel}) such that $\lambda$ is the spatial correlation length and $\sigma$ is the spatial localization length. In the subsequent analysis, scenarios $\textbf{S}_1$ and $\textbf{S}_2$ are treated simultaneously because, in both cases, the perturbation covariance is a Toeplitz operator since $\mathbf \Gamma(x,\xi)$ depends solely on the difference $x-\xi$ rather than the absolute locations $x$ and $\xi$. However, in scenario $\textbf{S}_3$, the perturbation covariance is spatially localized and $\mathbf \Gamma(x,\xi)$ depends on the absolute locations, and thus it is treated separately in subsection \ref{Section: Spatially Variant Covariance}. Recall that the linearized cochlear dynamics excludes micro-mechanical spatial coupling along different locations of the BM; whereas, scenario $\textbf{S}_2$ sort of reintroduces spatial coupling via the spatial correlations of the stochastic active gain. 

The condition of MSS (\ref{Eqn: MSS Condition}) can be rewritten as
\begin{equation} \label{Eqn: MSS Condition for S1, S2 and S3}
	\begin{aligned}
		\epsilon &< \frac{1}{\sqrt{\rho(\mathbb L_{\mathbf{\Gamma}})}},
	\end{aligned}
\end{equation}
for scenarios $\mathbf S_1, \mathbf S_2$ and $\mathbf S_3$. This bound is the maximum allowed perturbation in (\ref{Eqn: Stochastic Feedback}) such that MSS is maintained. In this section, we compute the upper bound on $\epsilon$ and the ``worst-case" covariance $\textbf{U}$ for the linearized cochlear dynamics. 

\subsection{Numerical Considerations}
This section describes the numerical considerations of the model and the numerical method used to compute the spectral radius and worst-case covariance of $\mathbb L_{\mathbf \Gamma}$. 

The numerical values of the parameters in this paper are taken from Table~I in \cite{ku2008statistics} for the linear cochlea. However, the expectation of the gain coefficient, $\bar \gamma(x)$, (which was considered to be spatially constant in \cite{ku2008statistics}) is left as a spatially distributed parameter to be tuned. The fluids block in Figure~\ref{Fig: Block Diagram of the Ear}(a) considered here is the one dimensional traveling wave as described in Appendix-\ref{Section: Mass Operators}. A spatial discretization grid of step size $\Delta_x := L/N_x$, where $N_x = 400$, is used to give a finite dimensional approximation of the operators (as matrices) describing the dynamics in Figure~\ref{Fig: Feedback Block Diagram} (refer to Appendix-\ref{Section: Finite Realizations}). 

Special care has to be taken when dealing with spatially white continuous processes (Scenario $\textbf{S}_1$). Let $\Gamma$ denote a matrix approximation of the uncertainty covariance operator $\mathbf \Gamma$ and approximate the Dirac delta function as 
\begin{equation} \label{Eqn: Delta Function}
	\begin{aligned}
		\delta(x) &\approx \frac{1}{\Delta_x} \text{rect}_{\Delta_x}(x) \\
		\text{such that}, \quad
		&\text{rect}_{\Delta_x} := 
		\begin{cases}
			1, & \text{if}\ -\frac{\Delta_x}{2} \leq x \leq \frac{\Delta_x}{2} \\
			0, & \text{otherwise}
		\end{cases}.
	\end{aligned}
\end{equation}	
Hence, the finite dimensional approximation of the perturbation covariance needs to be scaled with the discretization step $\Delta_x$ as follows 
\begin{equation} \label{Eqn: Discretized Covariance}
	\Gamma = \frac{1}{\Delta_x}I,
\end{equation}
where $I$ is the identity matrix.

Furthermore, our analysis requires the computation of the largest eigenvalue of the loop gain operator and its associated eigenvector (or eigen-operator). The matrices that approximate the spatial operators have a size of ($4(N_x+1) = 1604$), and keeping track of the underlying sparsity of all the approximated operators is essential for carrying out the computations efficiently. Note that to maintain the sparsity of (\ref{Eqn: Loop Gain Operator}) for scenario $\textbf{S}_2$, we use a truncated Gaussian kernel to approximate $\phi_{\lambda}$ given in (\ref{Eqn: Gaussian Kernel}), i.e. $ \phi_\lambda(x-\xi) \approx 0, \text{ for } |x-\xi| > d$, where $d$ is a pre-specified constant that represents a compromise between computational accuracy and sparsity. 
Finally, the power iteration method is employed for eigenvalue and eigenmatrix computations as recommended by \cite{parrilo2000cone}. This requires solving the Lyapunov-like equation in (\ref{Eqn: Loop Gain Operator}) at each iteration. 

\subsection{Stochastic Gain Coefficient with a Spatially Constant Expectation} \label{Section: Constant Expectation}
In this section, we set the expectation of the gain coefficient to one everywhere along the BM, i.e. $\bar \gamma(x) = 1$. To study the effects of the spatial correlations in the gain coefficient, we compare scenarios $\mathbf S_1$ and $\mathbf S_2$ by keeping in mind that $\mathbf S_1$ can be seen as a special case of $\mathbf S_2$ at the limit when $\lambda$ goes to zero. First, we compute the upper bounds on $\epsilon$ in (\ref{Eqn: MSS Condition for S1, S2 and S3}) such that MSS is maintained. Then we compute the worst-case covariance $\textbf{U}$ in (\ref{Eqn: BM Worst Case Covariance}). 

\begin{figure}[h]
	\centering
	\includegraphics[scale = .25]{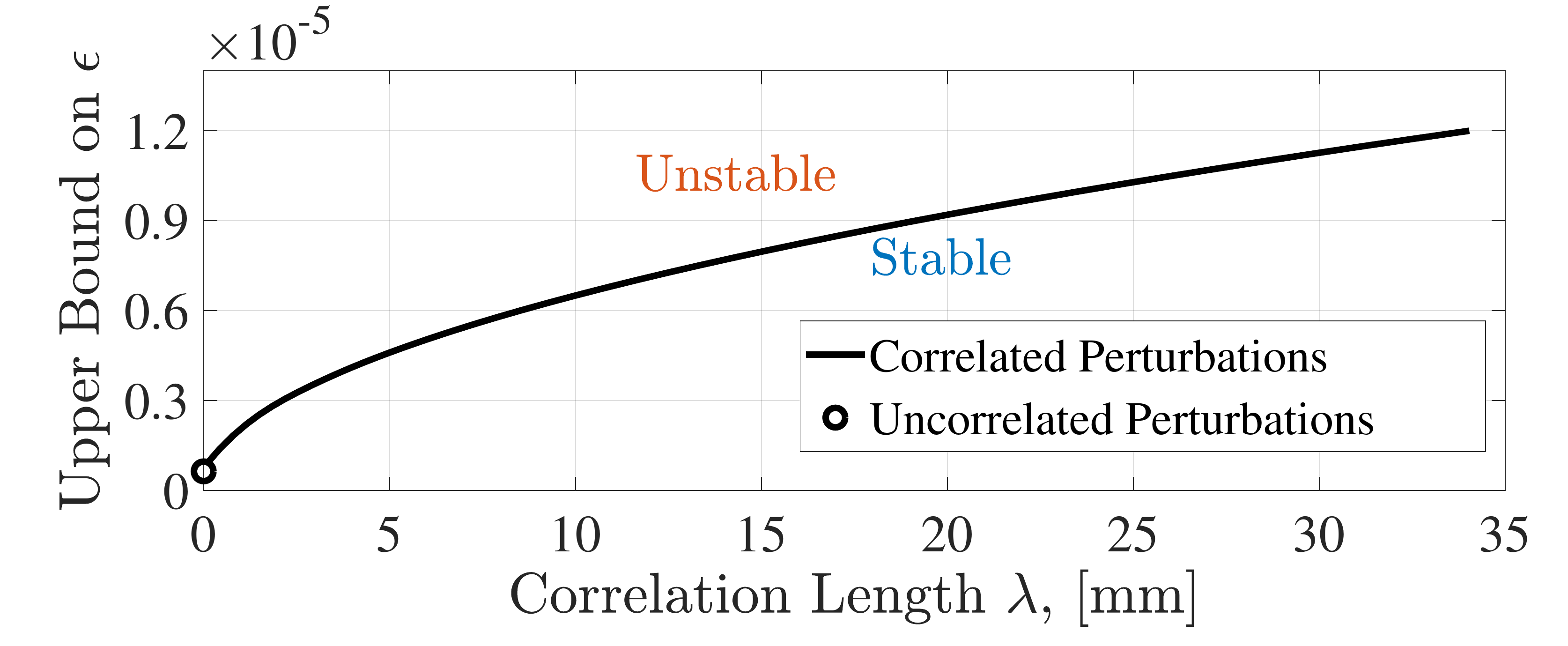}
	\caption{\footnotesize{Mean Square Stability Curve: Upper bound on the perturbation parameter, $\epsilon$, of the stochastic gain (\ref{Eqn: Stochastic Gain}) whose expectation is $\bar \gamma(x) = 1$. The black dot corresponds to scenario $\textbf{S}_1$ (uncorrelated gain perturbations) and the solid black line corresponds to scenario $\textbf{S}_2$ (correlated gain perturbations) for different spatial correlation lengths $\lambda$. The figure shows that larger correlation lengths make the model more immune to stochastic perturbations.}}
	\label{Fig: Sweeping lambda}
\end{figure}

By applying the power iteration method on (\ref{Eqn: Worst Case Covariance}), we compute the spectral radii $\rho(\mathbb L_{\mathbf \Gamma})$ and their associated eigen-operators $\mathbf P$ for scenarios $\textbf{S}_1$  and $\textbf{S}_2$  with different correlation lengths $\lambda$. Then, (\ref{Eqn: MSS Condition for S1, S2 and S3}) yields the upper bounds on $\epsilon$. The results are illustrated in Figure~\ref{Fig: Sweeping lambda} showing the small upper bounds on $\epsilon$. This reflects the high sensitivity of the model to such stochastic perturbations. As one would expect, a larger correlation length $\lambda$ requires a larger  perturbation to destabilize the linearized cochlea. 

\begin{figure}[h!]
	\centering
	\begin{tabular}{c}
		\includegraphics[scale = .25]{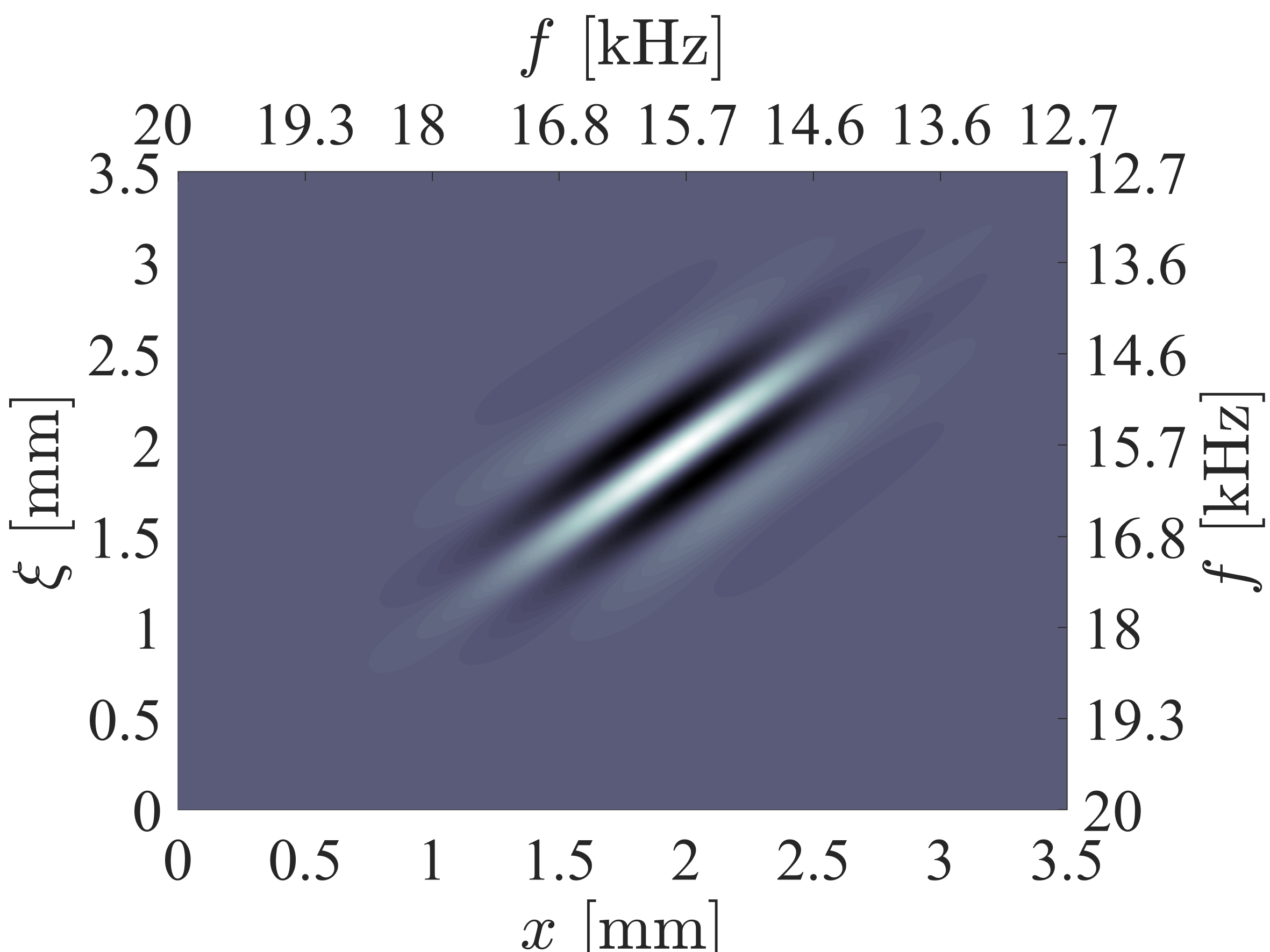} \\ \footnotesize{(a) Worst-Case Covariance of BM Displacement $\textbf{U}(x,\xi)$} \\ \\
		\includegraphics[scale = .25]{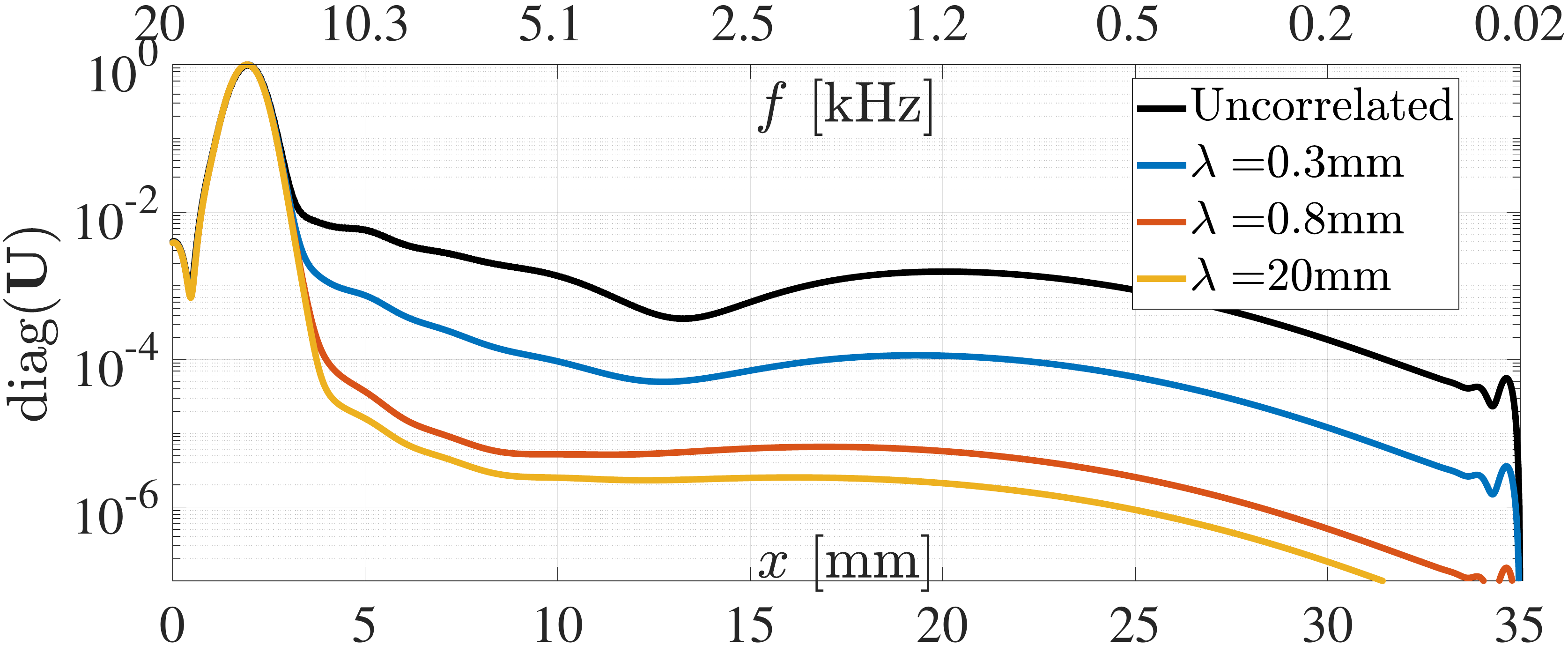} \\ \footnotesize{(b) Diagonal Entries of $\textbf{U}$}\\ \\
		\includegraphics[scale = .25]{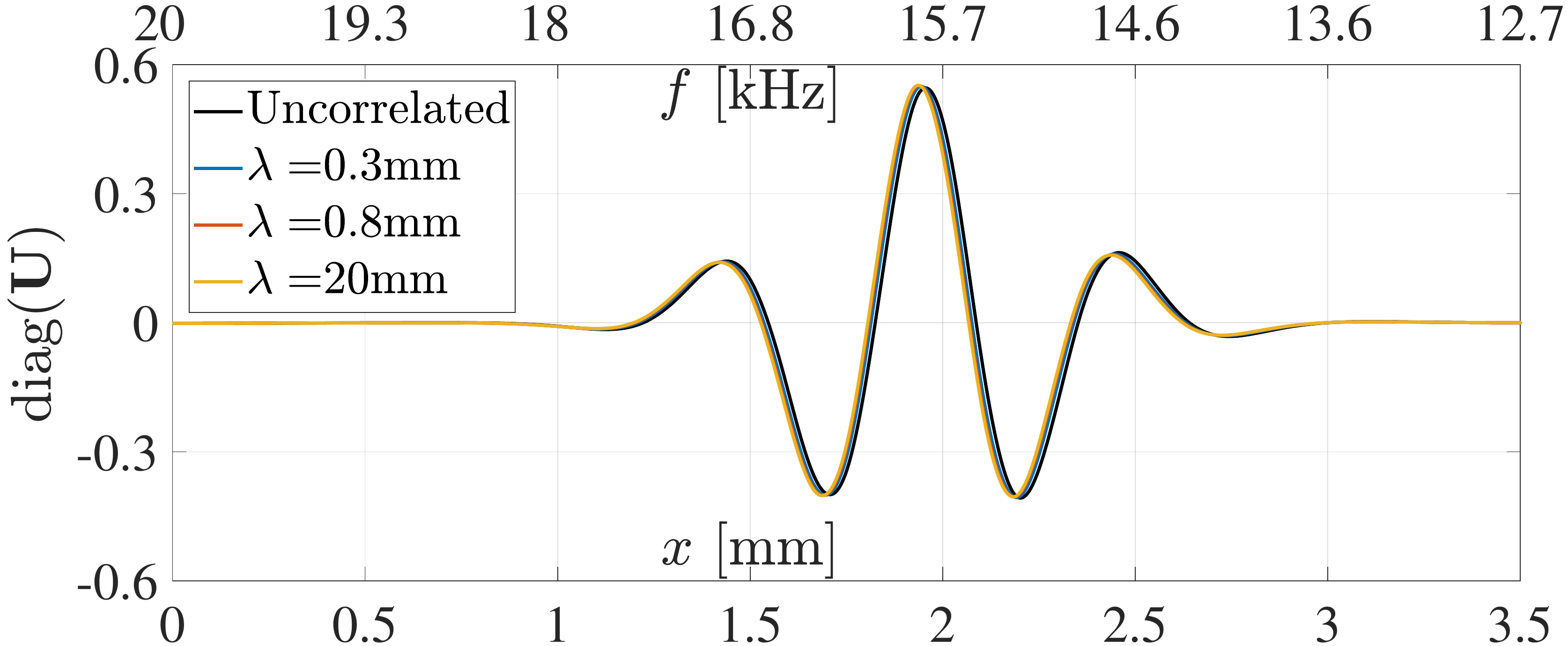} \\ \footnotesize{(c) Dominant Eigenfunction of $\textbf{U}$}
	\end{tabular}
	\caption{\footnotesize{Figure (a) shows an intensity plot of the worst-case covariance $\textbf{U}$ for scenario $\textbf{S}_1$ (uncorrelated gain perturbation) zoomed in for $0 \leq x,\xi \leq 3.5\si{mm}$. The axes correspond to the physical location $x$ in mm on the BM and the corresponding characteristic frequency $f$ in kHz. Figure (b) shows the diagonal entries of $\textbf{U}$ for scenarios $\textbf{S}_1$ and $\textbf{S}_2$ for different correlation lengths $\lambda$. Figure (c) depicts the dominant eigenfunction of $\textbf{U}$ for the different cases indicating the insignificant effect of $\lambda$ on the shape of the dominant eigenfunctions.}}
	\label{Fig: Sweeping lambda Covariance}
\end{figure}

The eigen-operator $\textbf{P}$ computed by the power iteration method is the worst-case pressure covariance. The corresponding worst-case covariance of the BM displacement $\textbf{U}$ is then computed using (\ref{Eqn: BM Worst Case Covariance}). Figure~\ref{Fig: Sweeping lambda Covariance}(a) shows $\textbf{U}$ for scenario $\textbf{S}_1$, zoomed in for $0 \leq x,\xi \leq L/10$. The intensity plot shows two sets of axes. The first axis represents the location on the BM and the second represents the corresponding characteristic frequency at each location, calculated using the Greenwood location-to-frequency mapping \cite{greenwood1990cochlear}. Observe that the covariance is band limited and the diagonal entries are dominant near the stapes ($x = 0$). This shows that instabilities essentially occur at high frequencies. Figure~\ref{Fig: Sweeping lambda Covariance}(b) plots the diagonal entries of $\textbf{U}$ for scenarios $\textbf{S}_1$ and $\textbf{S}_2$ for different correlation lengths $\lambda$. 
A smaller correlation length gives a slightly broader spectrum of unstable frequencies. However, for small $\epsilon$, the effect of the correlation length on the shape of the unstable BM modes is negligible. This is illustrated in Figure~\ref{Fig: Sweeping lambda Covariance}(c), where the dominant eigenfunction of $\textbf{U}$ is plotted for different cases.

\subsection{Stochastic Gain Coefficient with a Spatially Varying Expectation}
\begin{figure}[h]
	\centering
	\begin{tabular}{l}
		\includegraphics[scale = .25]{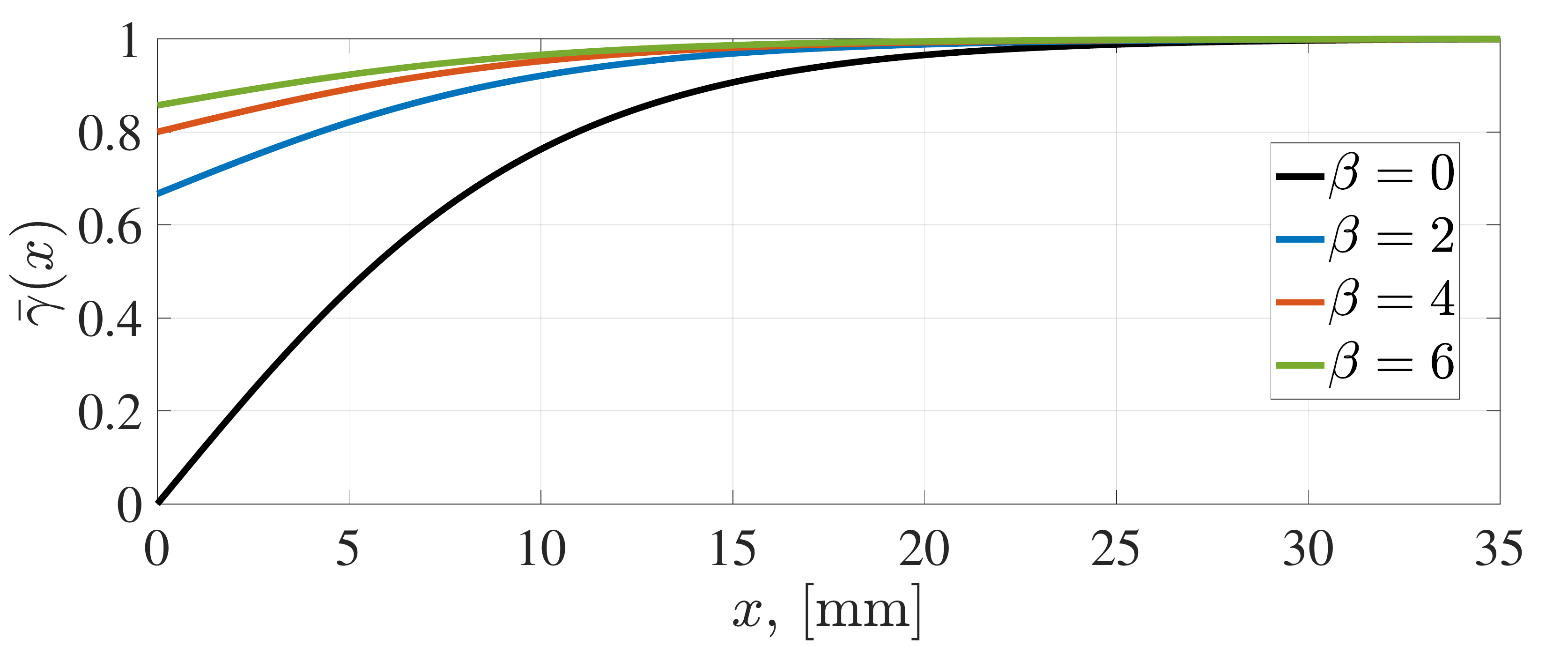} \\ \footnotesize{(a) Gain Coefficient Expectation Profiles} \\
		\includegraphics[scale = .25]{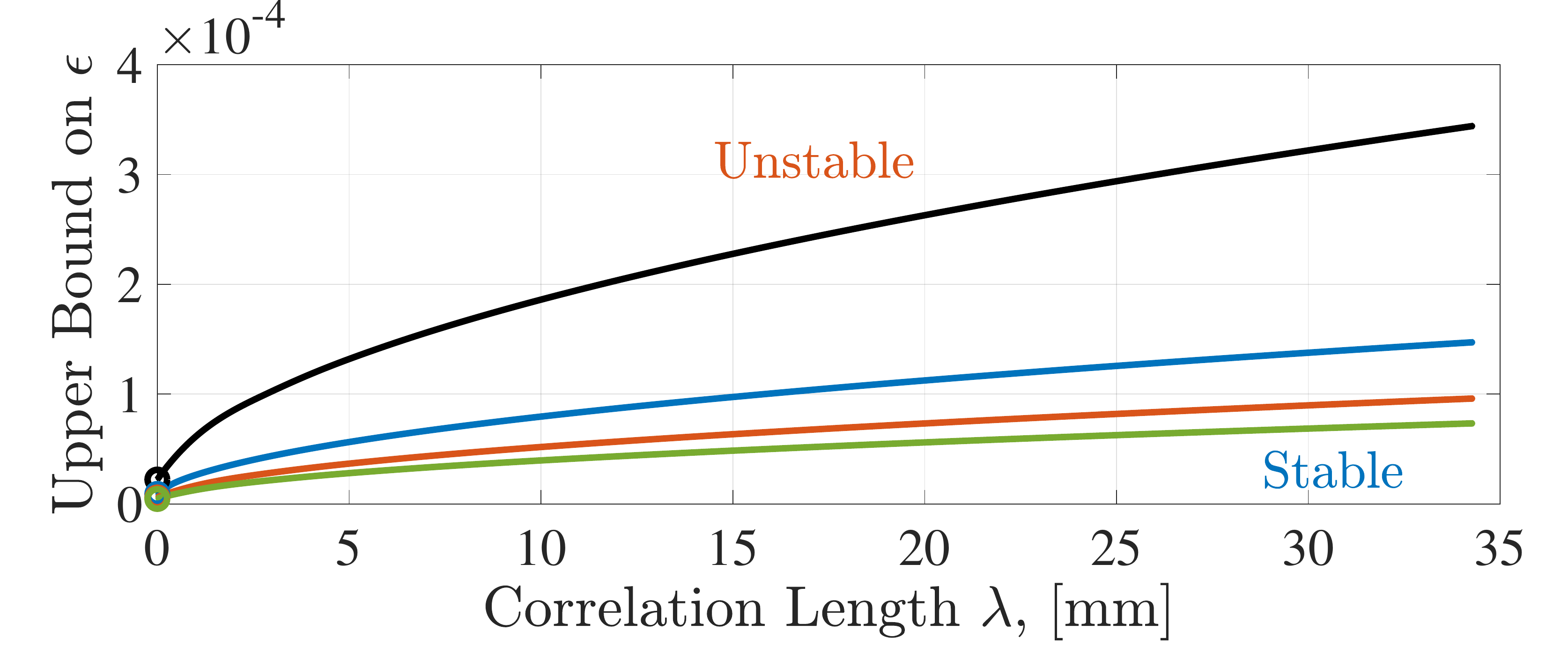}\\ \footnotesize{(b) Corresponding MSS Curves} \\
		\includegraphics[scale = .25]{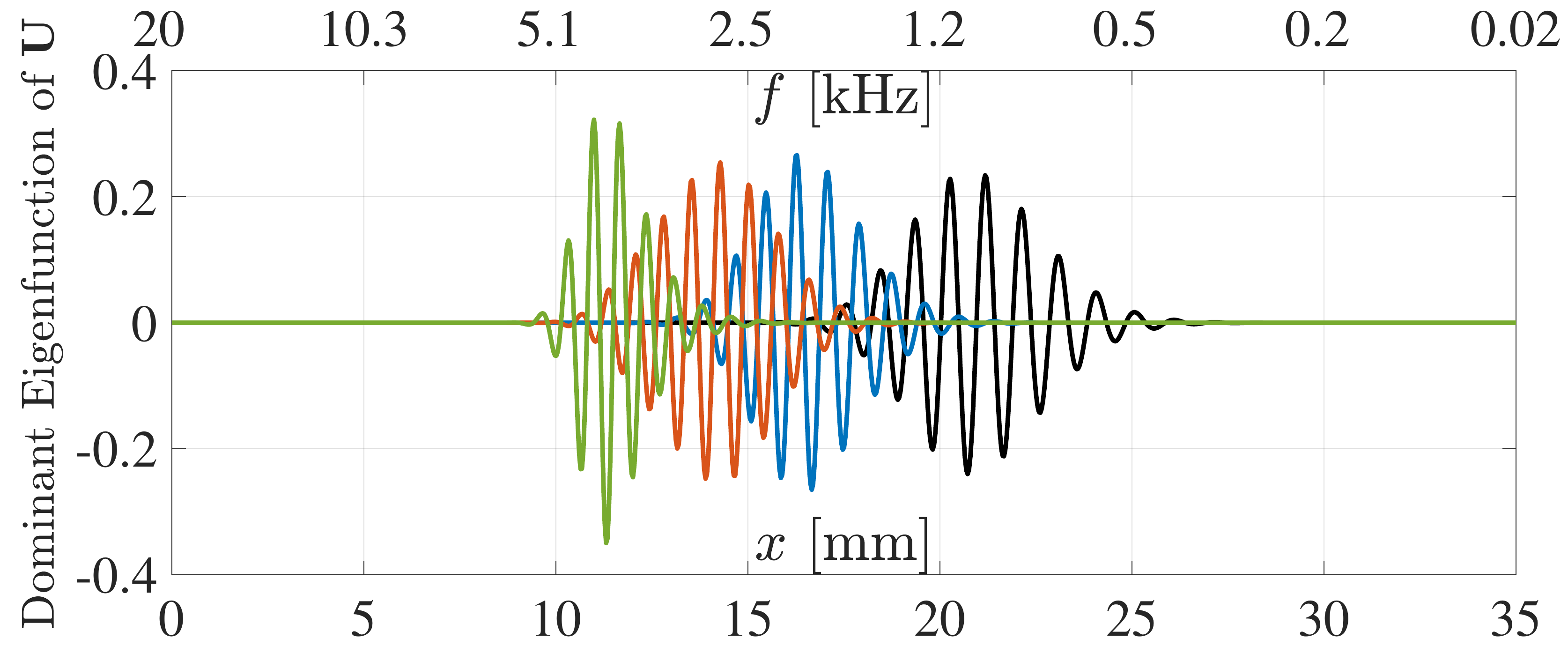}\\ \footnotesize{(c) Eigenfunctions for scenario $\mathbf S_1$} \\
	\end{tabular}
	\caption{\footnotesize{Mean Square Stability Curves for different gain coefficient expectation profiles: Figure(a) shows four different profiles of $\bar{\gamma}(x)$ generated as examples of spatially varying gain coefficients using (\ref{Eqn: Gain Mean Profiles}). The same values of $\beta$ are used in figures (b) and (c). Particularly, Figure(b) shows the upper bound on the perturbation parameter $\epsilon$ for the corresponding profiles of $\bar \gamma(x)$ in Figure(a). The circles correspond to scenario $\textbf{S}_1$ (uncorrelated gain perturbations) and the solid lines correspond to scenario $\textbf{S}_2$ (correlated gain perturbations) for different spatial correlation lengths $\lambda$. Figure (c) shows the eigenfunctions of the worst-case covariance operator $\textbf{U}$ corresponding to the different profiles of $\bar \gamma(x)$. The peaks of the eigenfunctions shift consistently with the shape of the gain profiles.}}
	\label{Fig: Sweeping lambda beta}
\end{figure}

This section shows that the frequencies of instabilities (or, equivalently, the locations on the BM) can shift depending on the shape of the expectation of the gain coefficient $\bar \gamma(x)$. For illustration purposes, four different profiles of $\bar{\gamma}_0(x)$ are generated as
\begin{equation} \label{Eqn: Gain Mean Profiles}
\bar{\gamma}_0(x) = \frac{\tanh(x/10) + \beta}{\tanh(L/10) + \beta},
\end{equation}
where $x$ and $L$ are expressed in mm and $\beta = 0,2,4$ and $6$. First, we show the MSS curves, similar to Figure~\ref{Fig: Sweeping lambda} for the four different profiles generated using (\ref{Eqn: Gain Mean Profiles}). Figure~\ref{Fig: Sweeping lambda beta}(b) clearly shows that the shape of $\bar \gamma(x)$ affects the margin of MSS. Particularly, the larger the dip in the gain coefficient, the higher $\epsilon$ needs to be to destabilize the linearized dynamics in the MSS sense. 

Since the correlation length for small values of $\epsilon$ has a negligible effect on the shape of the unstable modes as shown in Figure~\ref{Fig: Sweeping lambda Covariance}(c), we only present the worst-case covariances for scenario $\textbf{S}_1$. In fact, the correlation length only affects the margin of stability as illustrated in Figure~\ref{Fig: Sweeping lambda beta}(b). Figure~\ref{Fig: Sweeping lambda beta}(c) depicts the dominant eigenfunctions of $\textbf{U}$ for the four different profiles of $\bar \gamma(x)$. 
Clearly, the peaks of the unstable modes of the BM shift depending on the shape of $\bar \gamma(x)$. In fact, as the dip in $\bar \gamma(x)$ is increased, the peaks shift farther from the stapes resulting in instabilities of lower frequencies.

\subsection{Stochastic Gain Coefficient with a Spatially Localized Covariance} \label{Section: Spatially Variant Covariance}
We now treat the case where the gain coefficient $\gamma(x,t)$ in (\ref{Eqn: Stochastic Feedback}) has a spatially constant expectation, but spatially localized covariance given in scenario $\textbf{S}_3$, i.e.
$$ \bar \gamma(x) = 1 \qquad \text{and} \qquad \mathbf \Gamma(x,\xi) = \phi_\sigma(x-\mu) \delta(x-\xi),$$
for different values of $\sigma$ and $\mu$. Observe that for this form of $\mathbf \Gamma(x,\xi)$, the covariance is localized around $\mu$. Hence, this section investigates the cochlear instabilities that emerge as a result of stochastic perturbations localized around a particular location on the BM. 

\begin{figure}[h]
	\centering
	\begin{tabular}{c}
		\includegraphics[scale = .25]{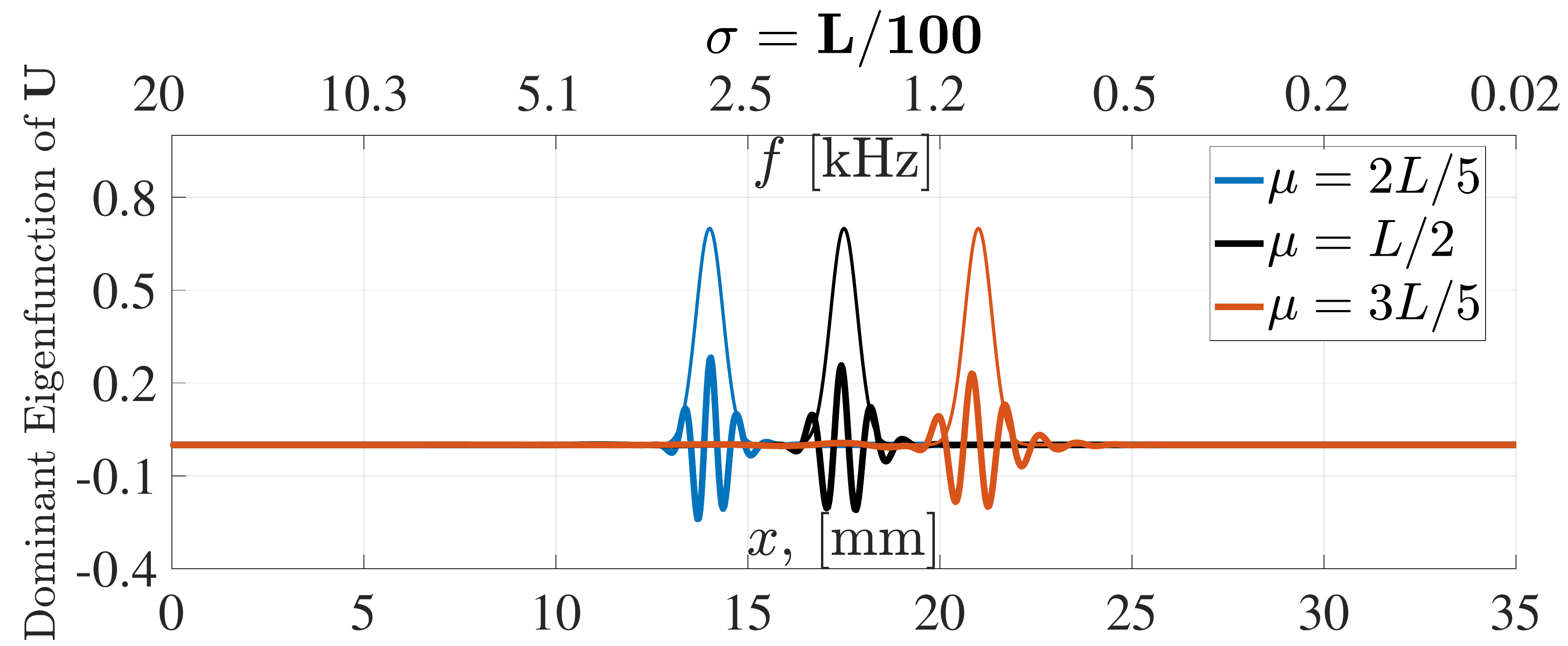} \\ (a) \\
		\includegraphics[scale = .25]{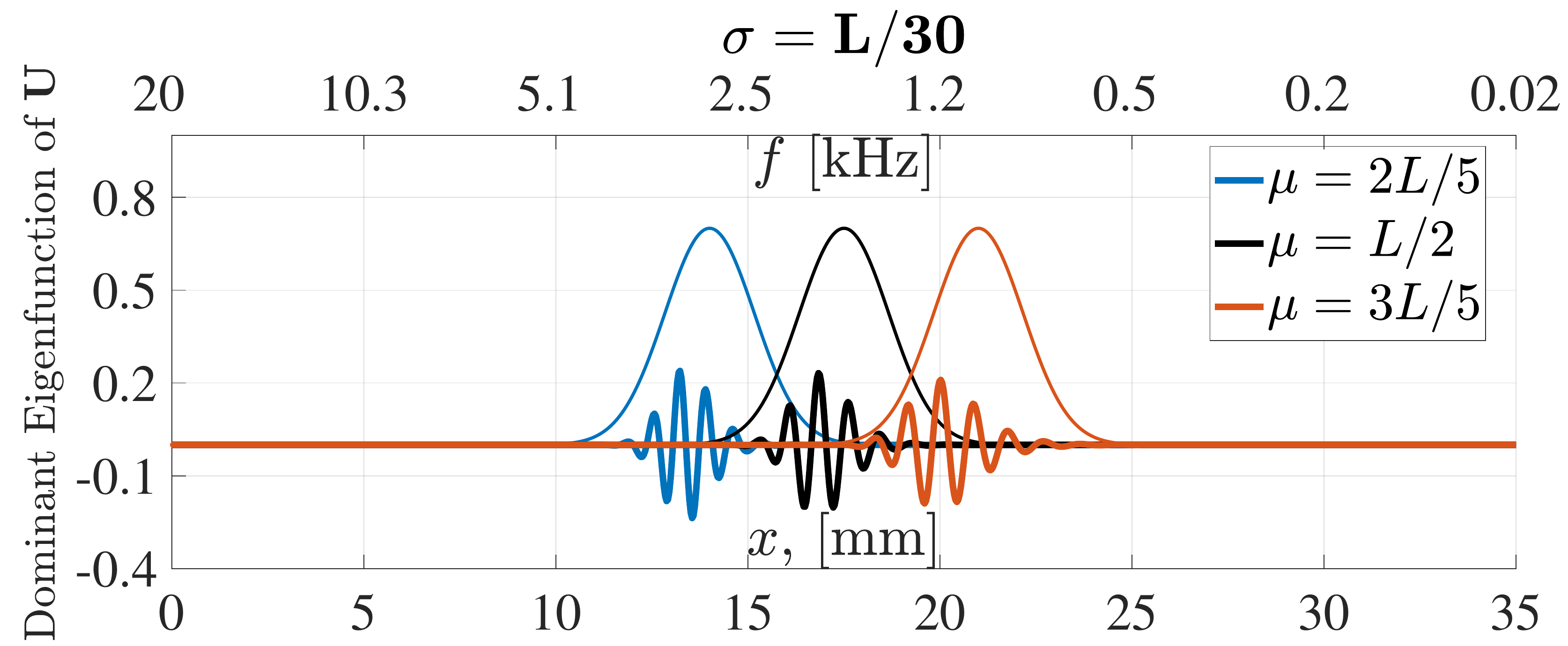} \\ (b) \\
		\includegraphics[scale = .25]{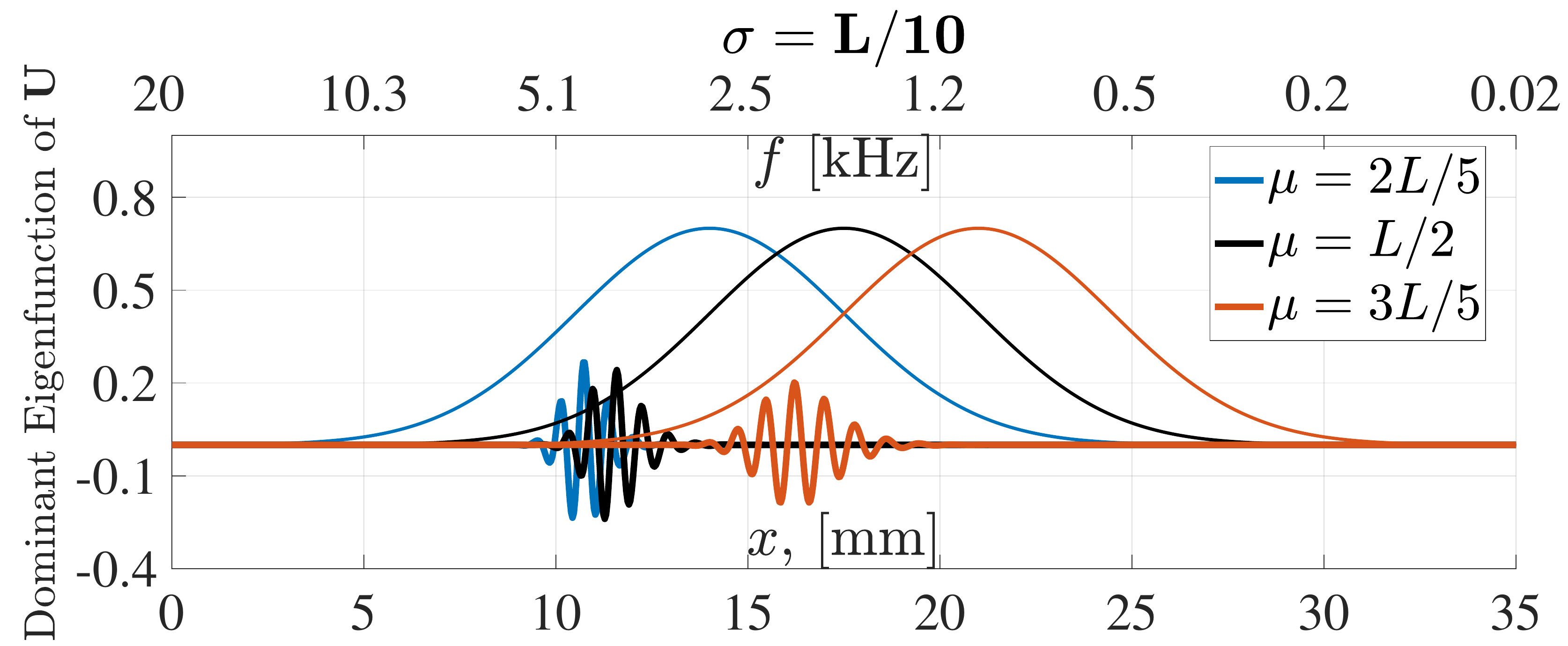} \\ (c) 
	\end{tabular}
	\caption{\footnotesize{Eigenfunctions of the worst-case covariance operator $\textbf{U}$ for different localized gain coefficient perturbations. These figures show the dominant eigenfunctions of the worst-case covariance operators for three different values of $\mu$ and $\sigma$. Particularly, in each figure, we fix $\sigma$ and vary $\mu$. Each thin curve represents a particular uncertainty spread function $\phi_{\sigma}(x-\mu)$ (not drawn to scale in the vertical axis) and each thick curve (with the same color) represents the corresponding dominant eigenfunction of the worst-case covariance operator. This figure illustrates the ``basal shifting" observation that resembles the phenomenon of detuning. }}
	\label{Fig: Spatially Localized Covariance}
\end{figure}

In particular, we are interested in tracking the unstable BM modes for different values of $\mu$ and $\sigma$, where $\mu$ is the location of the perturbation and $\sigma$ represents the local spread of the perturbation in the neighborhood of $\mu$. Following the same calculations of the previous sections, we compute the dominant eigenfunction of the worst-case covariance of the BM displacement $\mathbf U$ for different values of $\mu$ and $\sigma$.  The results are depicted in Figure~\ref{Fig: Spatially Localized Covariance}.
Observe that localized perturbations of the active gain coefficient at some location $\mu$ of the BM causes instabilities in that neighborhood. Particularly, for relatively small spread $\sigma = L/100$, the instabilities emerge at the same locations of the perturbations as shown in Figure~\ref{Fig: Spatially Localized Covariance}(a). However, as the spread of the uncertainty is increased up to $\sigma = L/30$ and $ L/10$, the location of the instability shifts towards the stapes. In fact, the wider the spread the larger the shift is as illustrated in Figures~\ref{Fig: Spatially Localized Covariance}(b) and (c).

This ``basal shifting" resembles the phenomenon of detuning observed in the cochlea. Acting as a frequency analyzer (or ``inverse-piano"), each location on the BM vibrates in response to a sound stimulus at a particular frequency. Thus, the BM has a frequency-to-location map such that every stimulus frequency has a preferred place on the BM called Characteristic Place (CP). The detuning phenomenon is observed as the shifting of the CP towards the stapes as the intensity of the stimulus (in dB) is increased. In this section, we showed that increasing the spread of the stochastic perturbations also shifts the BM vibrations towards the stapes. Nonlinear dynamics are necessary to model the detuning phenomenon. However, modeling this ``detuning-like" phenomenon doesn't require nonlinearities, instead a locally perturbed active gain is sufficient to explain it.

It is believed that these instabilities in the BM reflect back to the middle ear causing SOAEs \cite{nuttall2004spontaneous}. It is also believed that if these BM vibrations are intense enough, they can be perceived as tinnitus. Our results suggest a mechanism that explains the frequencies that can be detected in the ear canal due to SOAEs and/or perceived as tinnitus. As a matter of fact, the shape of the statistics (expectation and covariance) of the gain coefficient is a factor that controls the bands of the frequencies that are emitted as SOAEs. These emissions arise due to (a) \textit{spatially variant inhomogeneities} along the cochlear partition and (b) \textit{temporal stochastic perturbations} that give rise to structured stochastic uncertainties. 

\section{Nonlinear Stochastic Simulations} \label{Section: Simulations}
So far, the MSS analysis is carried out on the linearized dynamics. In this section, we carry out stochastic simulations of the nonlinear model to validate the predictions of our analysis of the linearized dynamics. 

\subsection{Nonlinear Descriptor State Space Formulation in Continuous Space-Time}
We first start by formulating the nonlinear dynamics in a DSS form similar to that given in section \ref{Section: DSS Formulation}. Recall that, the nonlinear deterministic active gain is given by (\ref{Eqn: Nonlinear Gain}) with $\gamma(x)$ representing the gain coefficient. To include stochastic perturbations, we substitute (\ref{Eqn: Stochastic Gain})  in (\ref{Eqn: Nonlinear Gain}) so that the nonlinear stochastic active gain can be written as
\begin{equation} \label{Eqn: Stochastic Nonlinear Active Gain}
	\begin{aligned}
		\left[\mathcal G(u)\right](x,t) &=\frac{\bar\gamma(x) + \epsilon \tilde \gamma(x,t)}{1 + \theta \left[\Phi_\eta\left(\frac{u^2}{R^2}\right)\right](x,t)}  \\
		& =: \bigg(\bar\gamma(x) + \epsilon \tilde \gamma(x,t)\bigg) \left[\tilde{\mathcal G}(u)\right](x,t).
	\end{aligned}
\end{equation}
Recall that $\Phi_{\eta}$ is the Gaussian spatial operator given by (\ref{Eqn: Guassian Weighing Operator}), $\theta = 0.5$, $R = 1\si{nm}$ and $\eta = 0.5345\si{mm}$. By substituting (\ref{Eqn: Stochastic Nonlinear Active Gain}) in (\ref{Eqn: Micromechanics}), we can rewrite the nonlinear model in a nonlinear DSS form as 
\begin{equation} \label{Eqn: Nonlinear DSS}
	\mathcal E \frac{\partial}{\partial t} \psi(x,t) = \big(\mathcal A_{\bar \gamma}(u) + \epsilon \tilde{\mathcal{B}}_0(u)\tilde \gamma \mathcal C_0\big) \psi(x,t),
\end{equation}
where $\mathcal A_{\bar{\gamma}}(u) := \mathcal A_0 + \tilde{\mathcal B_0}(u)  \bar \gamma \mathcal C_0$ and $\tilde{\mathcal{B}}_0(u)\tilde \gamma \mathcal C_0
$ are nonlinear spatial operators that represent the deterministic and stochastic portions of the dynamics, respectively. Note that $\mathcal E, \mathcal A_0$, and $\mathcal C_0$ are all defined in (\ref{Eqn: Descriptor State Space Operator Form}), and $\tilde{\mathcal B}_0(u) = \begin{bmatrix} 0 & 0 &  \tilde{\mathcal G}(u) & 0 \end{bmatrix}^T$. 
Therefore, (\ref{Eqn: Nonlinear DSS}) represents the nonlinear stochastic dynamics given in a DSS operator form, where the spatial variable is continuous. This is really a Stochastic Partial Differential Equation (SPDE) that needs to be discretized in space and time in order to carry out our simulations. 

\subsection{Description of the Numerical Method for Simulations}
In this section, we discretize (\ref{Eqn: Nonlinear DSS}) in space and time so that numerical simulations become fairly straightforward to implement. On a side note, if the stochastic perturbation $\tilde \gamma = 0$, (\ref{Eqn: Nonlinear DSS}) becomes a deterministic Partial Differential Equation (PDE). This can be easily integrated by discretizing space using a spatial grid, and then employ a time marching solver such as ODE45 in MATLAB. However, for an SPDE, one has to carefully treat the scaling of the covariances with the discretization steps. 

Space and time are discretized as $x_i = i\Delta_x$ and $t_n = n\Delta_t$ with discretization steps  $\Delta_x = L/N_x$ and $ \Delta_t = t_f/N_t$ for $i = 0, 1, ..., N_x$ and $ n = 0, 1, ..., N_t$, where $t_f$ is the final time. Let the BM and TM displacements on the discretized space-time grid be denoted by the vectors $u_n$ and $v_n \in \mathbb R^{N_x+1}$, respectively such that
\begin{align*}
	u_n &:= \begin{bmatrix} u(x_0,t_n) & \cdots & u(x_{N_x},t_n) \end{bmatrix}^T \\
	v_n &:= \begin{bmatrix} v(x_0,t_n) & \cdots & v(x_{N_x},t_n) \end{bmatrix}^T. 
\end{align*}
Then the discretized state space variable can be expressed by $\psi_n \in \mathbb R^{4(N_x+1)}$ as 
$$ \psi_n := \begin{bmatrix} u_n^T & v_n^T & \dot u_n^T & \dot v_n^T \end{bmatrix}^T.$$
For scenarios $\textbf{S}_1$ and $\textbf{S}_3$, $\tilde \gamma(x,t)$ is a zero-mean white process in space and time. It can be approximated at the spatial grid points $\{x_i\}_{i=0,1,...,N_x}$ and at time $t_n$ as follows
\begin{equation*}
	\begin{bmatrix} 
		\tilde \gamma(x_0,t_n) & \tilde \gamma(x_1,t_n) & \cdots & \tilde \gamma(x_{N_x},t_n)
	\end{bmatrix}^T\approx \frac{1}{\sqrt{\Delta_x \Delta_t}} w_n,
\end{equation*}
where $w_n \in \mathbb R^{N_x+1}$ is a zero-mean Gaussian random vector with a covariance matrix $\mathbb E\left[w_n w_n^T\right] = I$ for $\mathbf S_1$ and $\mathbb E\left[w_n w_n^T\right] = \mathcal D\left( \begin{bmatrix} \phi_{\sigma}(x_0-\mu) & \cdots & \phi_{\sigma}(x_{N_x}-\mu) \end{bmatrix}\right)$ for $\mathbf S_3$, where $\mathcal D$ is the diagonal operator such that $\mathcal D(w_n)$ is a diagonal matrix with $w_n$ arranged on its diagonal entries. 

For scenario $\textbf{S}_2$, $\tilde \gamma(x,t)$ is a stochastic process that is white in time but ``colored" in space with a spatial covariance $\mathbf \Gamma(x,\xi) = \epsilon^2 \phi_\lambda(x-\xi)$. In this scenario, the noise is smooth in space and there is no need to scale the covariance by the spatial discretization step. More precisely, $\tilde \gamma(x,t)$ can be approximated as
\begin{equation*}
	\begin{bmatrix} 
		\tilde \gamma(x_0,t_n) & \tilde \gamma(x_1,t_n) & \cdots & \tilde \gamma(x_{N_x},t_n)
	\end{bmatrix}^T\approx \frac{1}{\sqrt{\Delta_t}} w_n,
\end{equation*} 
where $\mathbb E\left[w_n w_n^T\right]$ is now  a symmetric matrix whose $(i,j)^{\text{th}}$ entry is given by $\phi_{\lambda}(x_i-x_j)$.

Therefore, a first order approximation of (\ref{Eqn: Nonlinear DSS}) can be carried out in the spirit of the Euler-Maruyama method \cite{maruyama1955continuous} to obtain
\begin{equation}\label{Eqn: Nonlinear Numerics}
	E \psi_{n+1} = E\psi_n + \Delta_t A_{\bar \gamma}(u_n) \psi_n + \alpha \tilde B_0(u_n) \mathcal D(w_n) C_0\psi_n
\end{equation}
where $\alpha = \epsilon \sqrt{\Delta_t/\Delta_x}$ for $\mathbf S_1$ and $\mathbf S_3$; and $\alpha = \epsilon \sqrt{\Delta_t}$ for $\mathbf S_2$.
 The matrices $E , A_{\bar{\gamma}}(u_n) , \tilde B_0(u_n)$ and $C_0 $ are all finite dimensional approximations of the operators $\mathcal E, \mathcal A_{\bar \gamma}(u), \tilde{\mathcal B}_0(u)$ and $\mathcal C_0$, respectively (Appendix-\ref{Section: Finite Realizations}).
Equation (\ref{Eqn: Nonlinear Numerics}) represents the recursive numerical methods to solve (\ref{Eqn: Nonlinear DSS}) for all three scenarios with the right choice of $\alpha$ and $\mathbb E[w_nw_n^T]$.

\subsection{Simulation of the Nonlinear Stochastic Model}
To validate our MSS analysis of the linearized dynamics and evaluate how well it copes with the nonlinear dynamics, we carry out a simulation of (\ref{Eqn: Nonlinear DSS}). This section considers scenario $\textbf{S}_1$. Hence, the numerical method used here is that given in (\ref{Eqn: Nonlinear Numerics}) for $\alpha = \epsilon^2 \sqrt{\Delta_t/\Delta_x}$ and $\mathbb E[w_nw_n^T] = I$.

\begin{figure}[h!]
	\centering
	\begin{tabular}{c}
		\includegraphics[scale = .22]{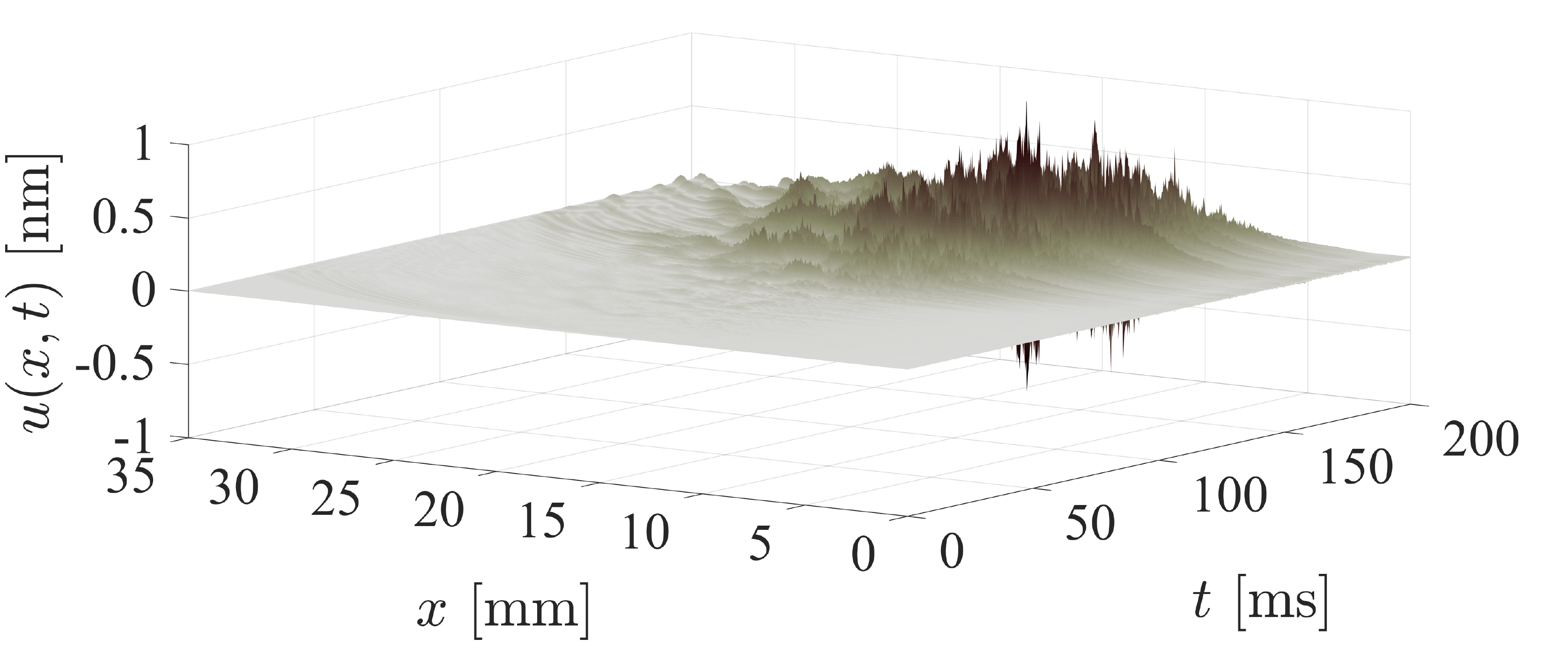} \\ \footnotesize{(a) Spatio-Temporal Stochastic Evolution of the BM} \\
		\includegraphics[scale = .22]{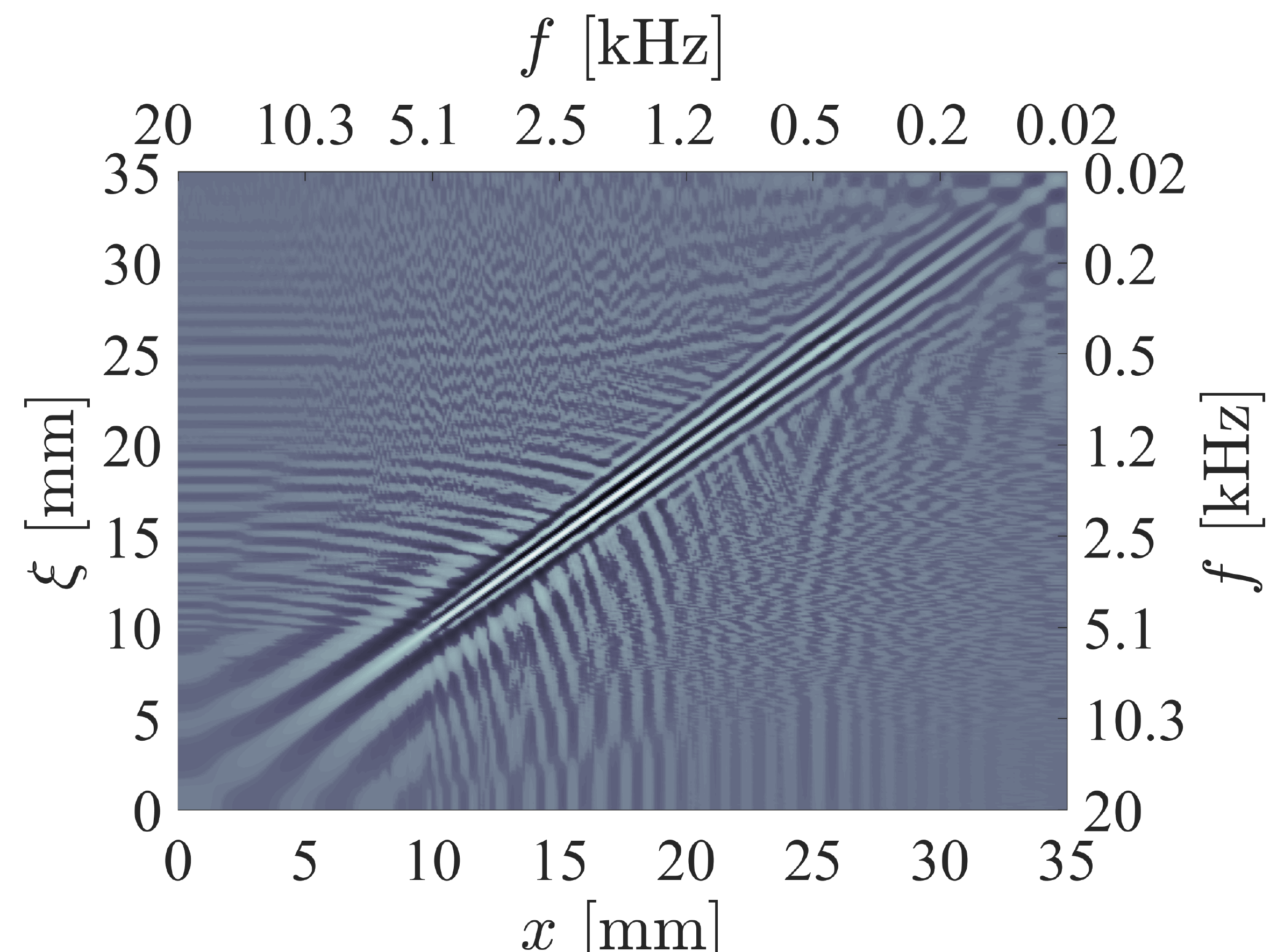} \\ \footnotesize{(b) Empirical Covariance $\mathbf U_{\text{Emp}}(x,\xi)$} \\
		\includegraphics[scale = .22]{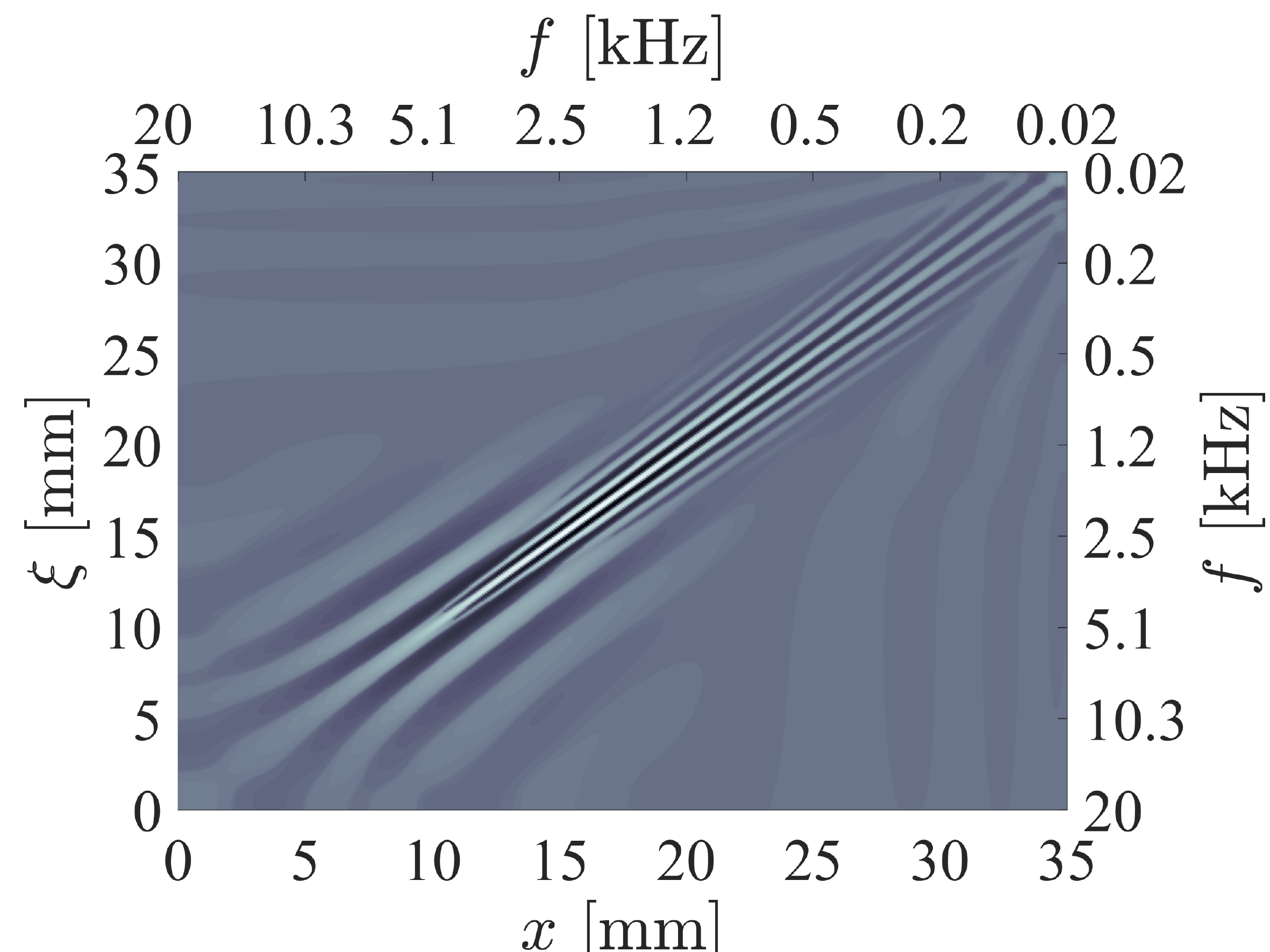} \\ \footnotesize{(c) Predicted Worst-Case Covariance $\mathbf U(x,\xi)$} \\
		\begin{tabular}{ll}
			\includegraphics[scale =.119]{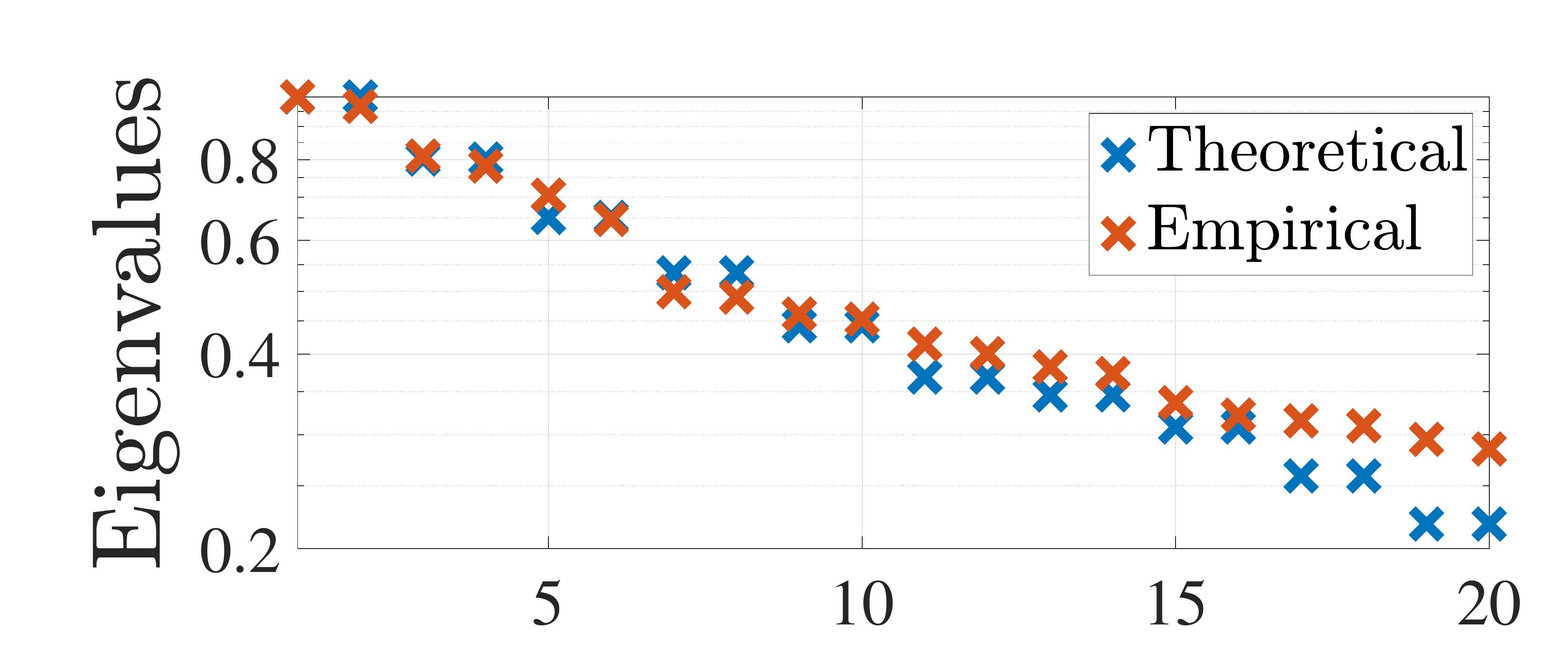} &
			\includegraphics[scale =.119]{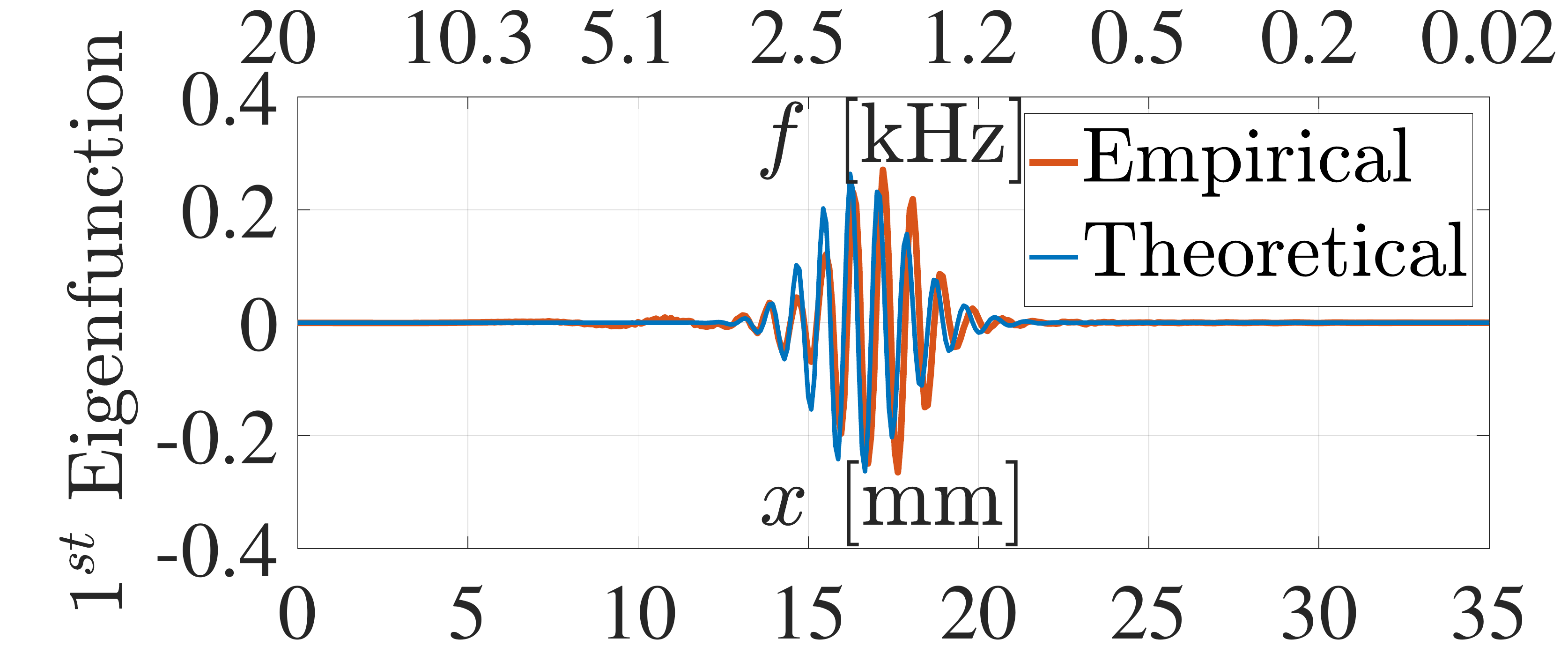} \\
			\includegraphics[scale=.119]{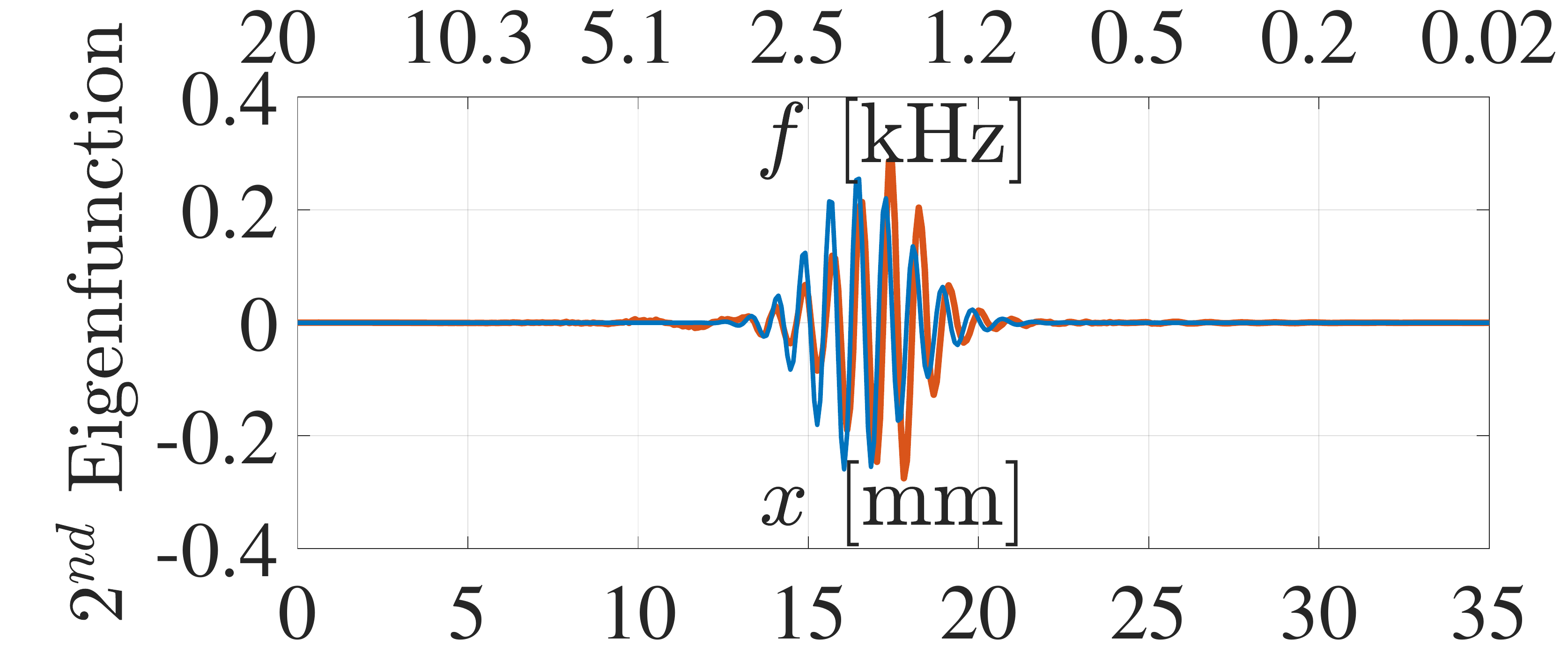}  & 
			\includegraphics[scale=.119]{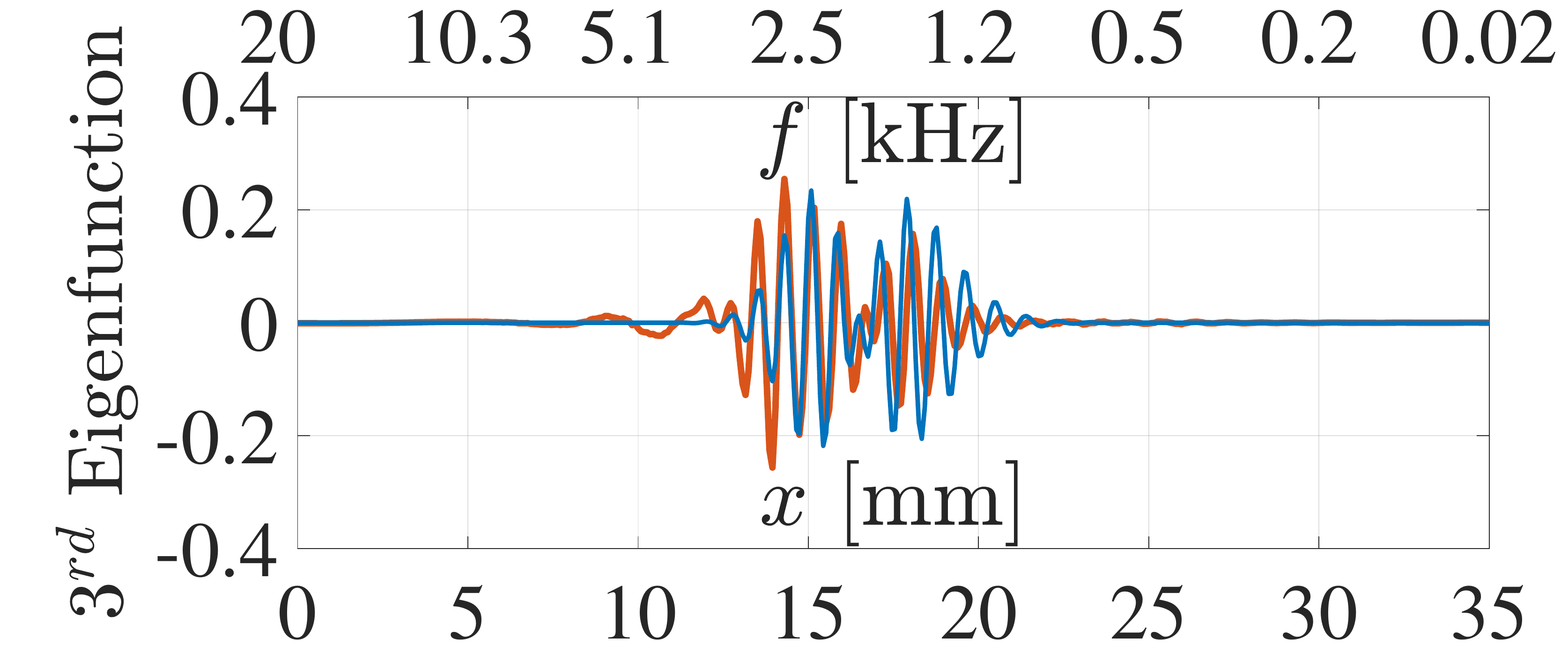} 
		\end{tabular}\\
		\footnotesize{(d) Empirical and Theoretical Dominant Eigenvalues/functions}
	\end{tabular}
	\caption{\footnotesize{Nonlinear Stochastic Simulation. Figure (a) shows the BM response to spatially uncorrelated stochastic active gain (scenario $\mathbf S_1$) with an expectation given by (\ref{Eqn: Gain Mean Profiles}) where $\beta = 2$ and a perturbation of $\epsilon = 1.1\times 10^{-5}$. Figures (b) and (c) show a comparison between the empirical and predicted covariances. The predicted covariance is computed for the linearized dynamics via the power iteration method applied on the loop gain operator (\ref{Eqn: Loop Gain Operator}). The empirical covariance is computed using the data obtained from one nonlinear stochastic simulation using (\ref{Eqn: Nonlinear Numerics}) and integrated in time using (\ref{Eqn: Ergodic Integration}) assuming ergodicity. Figure (d) shows a comparison between the dominant eigenvalues/functions of the empirical and predicted covariances shown in Figures (b) and (c), respectively.  This eigen-decomposition is referred to as the \textit{Karhunen{–}Lo\`ve} decomposition. Clearly the theoretical predictions match the empirical data, thus suggesting that the nonlinearities only saturate the response without significantly deforming the waveforms.}}
	\label{Fig: Nonlinear Stochastic Simulation}
\end{figure}

The nonlinear stochastic simulation shown here is for $\bar \gamma(x)$ given in (\ref{Eqn: Gain Mean Profiles}) with $\beta = 2$. All other scenarios are in agreement with our MSS analysis; however, this particular case study ($\beta = 2$) is chosen here to illustrate the effectiveness of our analysis. Observe using Figure~\ref{Fig: Sweeping lambda beta}(b) that for $\beta = 2$, the MSS condition is violated if $\epsilon \geq 9.1\times 10^{-6}$. We choose $\epsilon = 1.1 \times 10^{-5}$ which slightly violates the MSS condition for the linearized dynamics and allows the nonlinearity to kick in and saturate the response. The spatio-temporal response of the BM is depicted in Figure~\ref{Fig: Nonlinear Stochastic Simulation}(a) for $t \in [0, t_f]$ with $t_f = 200\si{ms}$. The response is maximal in a band limited region $10\si{mm} < x < 20\si{mm}$ which corresponds to a frequency range of roughly between $1\si{kHz}$ and $5\si{kHz}$. 
To be more precise, we compute the empirical covariance $\textbf{U}_{\text{Emp}}(x,\xi)$ as follows
\begin{equation} \label{Eqn: Ergodic Integration}
	\textbf{U}_{\text{Emp}}(x,\xi) = \frac{1}{t_f} \int_0^{t_f} u(x,\tau) u(\xi, \tau) d\tau.
\end{equation}
The time averaging replaces the expectation assuming ergodicity. Figures~\ref{Fig: Nonlinear Stochastic Simulation}(b) and (c) compare the empirical covariance to the predicted worst-case covariance. By visual inspection, we observe that the empirical results are in good agreement with our theoretical predictions. 
For a more precise comparison, we plot the first twenty dominant eigenvalues and first three dominant eigenfunctions of both the predicted and empirical covariances in Figure~\ref{Fig: Nonlinear Stochastic Simulation}(d). This eigen-decomposition is referred to as the \textit{Karhunen{-}Lo\`eve decomposition}. The eigenfunctions are the modes of BM vibrations that have the highest growth rate and are more likely to destabilize for small perturbations of the active gain. The plots doesn't show any significant difference between the empirical and theoretical results. In fact, although the nonlinear active gain slightly deforms the response, but its fundamental role (in the absence of a stimulus) is to saturate the linearized instabilities to form oscillations that remain bounded in time. 

\section{Discussion} \label{Section: Discussion}

The mechanisms underlying cochlear instabilities such as SOAEs and tinnitus are still controversial and not well understood. This paper suggests a new possible source of cochlear instabilities: spatio-temporal stochastic perturbations of the active gain. 

It is widely accepted that Outer Hair Cells (OHC) are responsible for the active gain in the cochlea. This work proposes a \textit{simulation-free}, system theoretic framework to analyze the effects of small stochastic perturbations that may occur on the level of the OHCs. These perturbations can have several physical origins such as noisy nearby neuronal activities, cellular activities, blood flow, etc... 

Studying the effects of randomness in the active gain is not new \cite{fruth2014active}, \cite{ku2008statistics}. However, the previous studies on this matter considered random spatial perturbations that are time-invariant. This type of randomness is referred to as ``frozen" or \textit{quenched disorder} in the statistical physics community. In fact, \cite{ku2008statistics} investigated the effects of the frozen spatial randomness by carrying out Monte Carlo simulations to study the statistics of the instabilities. However, to achieve a broad spectrum of unstable frequencies, the authors allowed severe perturbations of the active gain which is not realistic. Without these severe perturbations, the unstable frequencies would be limited to a band of high frequencies only (Section~\ref{Section: Constant Expectation}). This doesn't agree with the experimental observations where, for example, SOAEs are mainly found between $0.5$ and $4.5\si{kHz}$.  

A more realistic case is to treat the active gain as a stochastic process, where the randomness may occur in space and time, simultaneously. In addition to that, only small perturbations of the active gain are considered (three to four orders of magnitude less than \cite{ku2008statistics}). A major advantage of our analysis is that it is \textit{simulation-free} and no Monte Carlo simulations are required to study the statistics of the emerging instabilities. In our analysis, we also show that the band of unstable frequencies can be controlled by the tuning of the structural parameters of the cochlea such as the active gain coefficient. Hence, we show that even for very small perturbations, the unstable frequencies can be shifted dramatically. Furthermore, examining localized stochastic perturbations in the active gain allowed us to observe local instabilities that shift toward the stapes as the localization length or spread is larger. This observation resembles the detuning phenomenon present in the cochlea.

\section{Conclusion and Future Work}
This paper examines the instabilities that occur in the linearized dynamics due to spatio-temporal stochastic perturbations in the distributed structure of the cochlear partition. The simulation-free analysis is carried out through a structured stochastic uncertainty framework. It is shown that the spatial shape of the expectation and covariance of the gain coefficient affect the locations of the instabilities on the basilar membrane. These instabilities eventually saturate to form bounded oscillations due to the saturation nonlinearity of the active gain (\ref{Eqn: Nonlinear Gain}) producing spontaneous basilar membrane vibrations. It is believed that these instabilities are reflected to the middle ear as spontaneous otoacoustic emissions (SOAEs) \cite{nuttall2004spontaneous} with frequencies corresponding to the location of the instability on the basilar membrane. This analysis also suggests an explanation of one possible source of tinnitus, which is less addressed in the literature. Particularly, if the spontaneous BM vibrations were intense enough, they may be perceived as tinnitus. Future work will address instabilities that may occur due to stochastic uncertainties in structural parameters other than the active gain coefficient, such as the cochlear fluid density.


\bibliographystyle{ieeetr}
\bibliography{CochleaUncertaintyJASA_References}

\begin{thebibliography}{10}

\bibitem{fruth2014active}
F.~Fruth, F.~J{\"u}licher, and B.~Lindner, ``An active oscillator model
  describes the statistics of spontaneous otoacoustic emissions,'' {\em
  Biophysical journal}, vol.~107, no.~4, pp.~815--824, 2014.

\bibitem{ku2008statistics}
E.~M. Ku, S.~J. Elliott, and B.~Lineton, ``Statistics of instabilities in a
  state space model of the human cochlea,'' {\em The Journal of the Acoustical
  Society of America}, vol.~124, no.~2, pp.~1068--1079, 2008.

\bibitem{bamieh2018structured}
B.~Bamieh and M.~Filo, ``An input-output approach to structured stochastic
  uncertainty,'' {\em Submitted to IEEE Transactions on Automatic Control},
  2018.
\newblock Available online: \url{https://arxiv.org/abs/1806.07473}.

\bibitem{filo2018structured}
B.~Bamieh and M.~Filo, ``An input-output approach to structured stochastic
  uncertainty in continuous time,'' {\em Submitted to IEEE Transactions on
  Automatic Control}, 2018.
\newblock Available online: \url{https://arxiv.org/abs/1806.09091}.

\bibitem{lu2002mean}
J.~Lu and R.~E. Skelton, ``Mean-square small gain theorem for stochastic
  control: discrete-time case,'' {\em IEEE Transactions on Automatic Control},
  vol.~47, no.~3, pp.~490--494, 2002.

\bibitem{elia2005remote}
N.~Elia, ``Remote stabilization over fading channels,'' {\em Systems \& Control
  Letters}, vol.~54, no.~3, pp.~237--249, 2005.

\bibitem{geisler1995cochlear}
C.~D. Geisler and C.~Sang, ``A cochlear model using feed-forward
  outer-hair-cell forces,'' {\em Hearing research}, vol.~86, no.~1,
  pp.~132--146, 1995.

\bibitem{lamar2006signal}
M.~D. LaMar, J.~Xin, and Y.~Qi, ``Signal processing of acoustic signals in the
  time domain with an active nonlinear nonlocal cochlear model,'' {\em Signal
  processing}, vol.~86, no.~2, pp.~360--374, 2006.

\bibitem{neely1986model}
S.~T. Neely and D.~Kim, ``A model for active elements in cochlear
  biomechanics,'' {\em The Journal of the Acoustical Society of America},
  vol.~79, no.~5, pp.~1472--1480, 1986.

\bibitem{steele1979comparison}
C.~R. Steele and L.~A. Taber, ``Comparison of wkb and finite difference
  calculations for a two-dimensional cochlear model,'' {\em The Journal of the
  Acoustical Society of America}, vol.~65, no.~4, pp.~1001--1006, 1979.

\bibitem{givelberg2003comprehensive}
E.~Givelberg and J.~Bunn, ``A comprehensive three-dimensional model of the
  cochlea,'' {\em Journal of Computational Physics}, vol.~191, no.~2,
  pp.~377--391, 2003.

\bibitem{neely1981finite}
S.~T. Neely, ``Finite difference solution of a two-dimensional mathematical
  model of the cochlea,'' {\em The Journal of the Acoustical Society of
  America}, vol.~69, no.~5, pp.~1386--1393, 1981.

\bibitem{elliott2013wave}
S.~J. Elliott, G.~Ni, B.~R. Mace, and B.~Lineton, ``A wave finite element
  analysis of the passive cochlea,'' {\em The Journal of the Acoustical Society
  of America}, vol.~133, no.~3, pp.~1535--1545, 2013.

\bibitem{elliott2007state}
S.~J. Elliott, E.~M. Ku, and B.~Lineton, ``A state space model for cochlear
  mechanics,'' {\em The Journal of the Acoustical Society of America},
  vol.~122, no.~5, pp.~2759--2771, 2007.

\bibitem{bertaccini2011fast}
D.~Bertaccini and R.~Sisto, ``Fast numerical solution of nonlinear nonlocal
  cochlear models,'' {\em Journal of Computational Physics}, vol.~230, no.~7,
  pp.~2575--2587, 2011.

\bibitem{filo2016order}
M.~Filo, F.~Karameh, and M.~Awad, ``Order reduction and efficient
  implementation of nonlinear nonlocal cochlear response models,'' {\em
  Biological cybernetics}, vol.~110, no.~6, pp.~435--454, 2016.

\bibitem{filo2017topics}
M.~G. Filo, {\em Topics in Modeling of Cochlear Dynamics: Computation, Response
  and Stability Analysis}.
\newblock PhD thesis, University of California, Santa Barbara, 2017.

\bibitem{zhou1996robust}
K.~Zhou, J.~C. Doyle, K.~Glover, {\em et~al.}, {\em Robust and optimal
  control}, vol.~40.
\newblock Prentice hall New Jersey, 1996.

\bibitem{oksendal2003stochastic}
B.~{\O}ksendal, ``Stochastic differential equations,'' in {\em Stochastic
  differential equations}, pp.~65--84, Springer, 2003.

\bibitem{parrilo2000cone}
P.~A. Parrilo and S.~Khatri, ``On cone-invariant linear matrix inequalities,''
  {\em IEEE Transactions on Automatic Control}, vol.~45, no.~8, pp.~1558--1563,
  2000.

\bibitem{greenwood1990cochlear}
D.~D. Greenwood, ``A cochlear frequency-position function for several
  species—29 years later,'' {\em The Journal of the Acoustical Society of
  America}, vol.~87, no.~6, pp.~2592--2605, 1990.

\bibitem{nuttall2004spontaneous}
A.~Nuttall, K.~Grosh, J.~Zheng, E.~De~Boer, Y.~Zou, and T.~Ren, ``Spontaneous
  basilar membrane oscillation and otoacoustic emission at 15 khz in a guinea
  pig,'' {\em Journal of the Association for Research in Otolaryngology},
  vol.~5, no.~4, pp.~337--348, 2004.

\bibitem{maruyama1955continuous}
G.~Maruyama, ``Continuous markov processes and stochastic equations,'' {\em
  Rendiconti del Circolo Matematico di Palermo}, vol.~4, no.~1, p.~48, 1955.

\bibitem{moore1986parallels}
B.~Moore, ``Parallels between frequency selectivity measured psychophysically
  and in cochlear mechanics,'' 1986.

\end{thebibliography}

\appendix
\subsection{Mass Operators} \label{Section: Mass Operators}
The fluids block in Figure~\ref{Fig: Block Diagram of the Ear}(a) can be modeled in 1D, 2D, or 3D. In two dimensions, Navier-Stokes equations boil down to the Laplace equation with the appropriate boundary conditions as shown in \cite{lamar2006signal}. This simplification is valid under the assumptions of incompressible, inviscid fluid where the magnitude of the vibrations of the membranes are negligible relative to the dimensions of the cochlea. These assumptions make the fluid block in Figure~\ref{Fig: Block Diagram of the Ear} memoryless and amenable to be represented by the two linear spatial operators $\mathcal M_f$ and $\mathcal M_s$ in (\ref{Eqn: BM Pressure}). In this paper, we give these operators for the 1D case only. Higher dimensions can be treated similarly. As in \cite{elliott2007state}, the fluid block in 1D can be represented by the traveling wave equation as follows
\begin{equation} \label{Eqn: PDE for 1D Fluid}
\begin{aligned}
\frac{\partial^2}{\partial x^2} p(x,t) &= \frac{2\rho}{H}\ddot u(x,t); ~
\begin{cases}
\frac{\partial}{\partial x} p(0,t) = 2\rho \ddot s(t) \\
p(L,t) = 0,
\end{cases}
\end{aligned}
\end{equation}
where $\rho$ is the density of the fluid, $H$ is the height of the fluid chamber and $L$ is the length of the BM. This is a linear system with two inputs: $\ddot u$ and $\ddot s$. It can be shown that the solution of (\ref{Eqn: PDE for 1D Fluid}) is
\begin{equation} \label{Eqn: Mass Operators Expression}
\begin{aligned}
p(x,t) &= -[\mathcal M_f \ddot u](x,t) - [\mathcal M_s] \ddot s(t) \\ 
[\mathcal M_f \ddot u](x,t) &:= -\frac{2\rho}{H} \sum_{n=0}^{\infty} \frac{1}{\lambda_n} \phi_n(x) \langle \phi_n, \ddot u(.,t) \rangle \\
[\mathcal M_s \ddot s](t) &:= 2\rho (L-x) \ddot s(t),
\end{aligned}
\end{equation}
where $\langle ., .\rangle$ denotes the inner product in the space of square integrable functions over $[0,L]$, and
\begin{equation*}
\begin{aligned}
\lambda_n=-(n+\frac{1}{2})^2\frac{\pi^2}{L^2} \longleftrightarrow \phi_n(x) =
\sqrt{\frac{2}{L}}\cos\left[\left(n+\frac{1}{2}\right)\frac{\pi}{L}x\right], \\
\end{aligned}
\end{equation*}
for $n=0,1,2,...$ 
It is fairly straightforward to verify that (\ref{Eqn: Mass Operators Expression}) is indeed a solution by substituting in (\ref{Eqn: PDE for 1D Fluid}). 

Finite dimensional approximations can be obtained by representing $\mathcal M_f$ and $\mathcal M_s$ by the matrix $M_f \in \mathbb R^{(N_x+1) \times (N_x+1)}$ and the vector $M_s \in \mathbb R^{N_x+1}$, respectively, where $N_x+1$ is the spatial grid size that discretizes the spatial variable $x$. This is done by truncating the sum and by using a quadrature rule to compute the inner product (or simply a trapezoidal rule). Note that finite difference methods, in the spirit of \cite{elliott2007state} and \cite{neely1981finite}, can also be used to approximate the mass operators. However, the spectral method we presented here provides a better and more efficient approximation. 

\subsection{Matrix Approximation of Spatial Operators} \label{Section: Finite Realizations}
Let the matrices 
\begin{align*}
&F_\eta \in \mathbb R^{(N_x+1)\times (N_x+1)};  &A_0 \in \mathbb R^{4(N_x+1)\times 4(N_x+1)}; \\
&B_0 \in \mathbb R^{4(N_x+1)\times (N_x+1)}; &  \tilde B_0(u_n) \in \mathbb R^{4(N_x+1)\times (N_x+1)};\\
&C_0 \in \mathbb R^{(N_x+1)\times 4(N_x+1)};  & E \in \mathbb R^{4(N_x+1)\times 4(N_x+1)}; \\
&A_{\bar{\gamma}}(u_n) \in \mathbb R^{4(N_x+1)\times 4(N_x+1)},
\end{align*}
be the finite dimensional approximations of the spatial operators $\Phi_\eta, \mathcal A_0, \mathcal B_0, \tilde{\mathcal B}_0(u), \mathcal C_0, \mathcal E$ and $ \mathcal A_{\bar{\gamma}}(u_n)$, respectively. 
Using the trapezoidal integration rule on (\ref{Eqn: Guassian Weighing Operator}), we can construct the matrix $F_\eta$ as
$$F_\eta = \mathcal D\left(\tilde F_\eta T \boldsymbol{1}\right)^{-1} \tilde F_\eta T,$$
where $\mathcal D$ is the diagonal operator, $\tilde F_\eta\in \mathbb R^{(N_x+1)\times(N_x+1)}$ and its $(i,j)^{th}$ entry is defined as $\left(\tilde F_\eta\right)_{ij} := e^{-(i-j)^2\frac{\Delta_x^2}{\eta^2}}$, $\mathbf{1} \in \mathbb R^{N_x+1}$ is a vector whose entries are all ones and $T \in \mathbb R^{(N_x+1)\times(N_x+1)}$ is a diagonal matrix defined as 
$$ T := \mathcal D\left(\begin{bmatrix} \frac{1}{2} & 1 & \cdots & 1 & \frac{1}{2} \end{bmatrix}\right).$$
Furthermore, define the following diagonal matrices $\in \mathbb R^{(N_x+1)\times(N_x+1)}$
\begin{align*}
K_l &:= \mathcal D\left(\begin{bmatrix} k_l(x_0) & \cdots & k_l(x_{N_x}) \end{bmatrix}\right), \quad l = 1, 2, 3, 4;\\
C_l &:= \mathcal D\left(\begin{bmatrix} c_l(x_0) & \cdots & c_l(x_{N_x}) \end{bmatrix}\right), \quad l = 1, 2, 3, 4;\\
D_{\bar \gamma} &:= \mathcal D\left(\begin{bmatrix} \bar \gamma(x_0) & \cdots & \bar \gamma(x_{N_x}) \end{bmatrix}\right); \\ 
\tilde G(u_n) &:= \mathcal D\left(1+\frac{\theta}{R^2} F_\eta (u_n \circ u_n)\right)^{-1} ,
\end{align*}
where $\circ$ is the Hadamard (element-by-element) product.
Therefore
\begin{align*}
& E := 
\begin{bmatrix}
I 		& 0				& 0											& 0					\\
0				& I	& 0											& 0					\\
0				& 0				& \frac{g}{b}m_1 I  + M_f	& 0					\\
0				& 0				& 0											& m_2 I 	\\
\end{bmatrix}; \quad B := 
\begin{bmatrix} 0 \\ 0 \\ -M_s \\ 0 \end{bmatrix}; \\
& \Scale[0.98]{A_0 := \resizebox{0.92\hsize}{!}{$
	\begin{bmatrix}
	0						& 0				& I				& 0\\
	0						& 0				& 0							& I\\
	-\frac{g}{b}(K_1+K_3)	& K_3			& -\frac{g}{b}(C_1+C_3)		& C_3\\
	\frac{g}{b}K_3			& -(K_2+K_3)	& \frac{g}{b}C_3			&-(C_2+C_3)	
	\end{bmatrix}$};}\\
& 	B_0 := \begin{bmatrix} 0 & 0 &  I & 0 \end{bmatrix}^T ;\quad 
C_0 := \begin{bmatrix} \frac{g}{b}K_4 & -K_4 & \frac{g}{b}C_4 & -C_4 \end{bmatrix} ;\\
& 	\tilde B_0(u_n) := \begin{bmatrix} 0 & 0 &  \tilde G(u_n) & 0 \end{bmatrix}^T; \\
&A_{\bar{\gamma}}(u_n) := A_0 + \tilde B_0(u_n) D_{\bar \gamma} C_0.
\end{align*}

\subsection{System Linearization} \label{Section: System Linearization}
The only nonlinear portion of the dynamics appears in the active gain given by (\ref{Eqn: Nonlinear Gain}). Thus, to linearize the dynamics around the origin, it suffices to linearize the active gain. Up to first order, the active gain can be expanded around some $\bar u$, by letting ${u := \bar u + \epsilon \tilde u}$. The expansion is given by
$$\mathcal G(u) = \mathcal G(\bar u) + \epsilon \left[\frac{\partial}{\partial u} \mathcal G(\bar u)\right](\tilde u) + \mathcal O(\epsilon^2), $$
where $\left[\frac{\partial}{\partial u} \mathcal G(\bar u)\right](\tilde u)$ is the Fr\'echet derivative in the direction of $\tilde u$. It can be calculated as follows
\begin{align*}
\left[\frac{\partial}{\partial u} \mathcal G(\bar u)\right](\tilde u) &:= \lim_{\epsilon\to 0} \frac{\mathcal G(\bar u + \epsilon \tilde u) - \mathcal G(\bar u)}{\epsilon} \\
&= -\frac{2\theta}{R^2}\frac{\gamma \Phi_\eta(\bar u \tilde u)}{\left(1 + \theta\Phi_\eta\left(\frac{u^2}{R^2}\right)\right)^2}.
\end{align*}
To linearize around the origin, we set $\bar u = 0$. This yields
\begin{align*}
\mathcal G(0) = \gamma \qquad \text{and} \qquad \left[\frac{\partial}{\partial u} \mathcal G(0)\right](\tilde u) = 0.
\end{align*}
Therefore, up to first order, the linearization around the fixed point of the active gain is $ \mathcal G(u) = \gamma + \mathcal O (\epsilon^2).$

\subsection{Equivalent Rectangular Bandwidth} \label{Section: ERB}
The width, $\eta$, of the Gaussian kernel in (\ref{Eqn: Gaussian Kernel}) controls the spatial coupling length along the BM. The numerical value of $\eta$ in this paper is chosen based on the critical bands
in the cochlea. In psychoacoustics, the concept of critical bands was introduced by Harvey Fletcher in 1933. He described the bands of audio frequencies within which two tones interfere in the perception of each other, thus indicating the length of spatial coupling along the cochlea. This band, which is termed Equivalent Rectangular Bandwidth (ERB), is believed to be equivalent to $0.89 \si{mm}$ on the BM \cite{moore1986parallels}. 

\begin{figure}[h!]
	\centering
	\includegraphics[scale = 0.32]{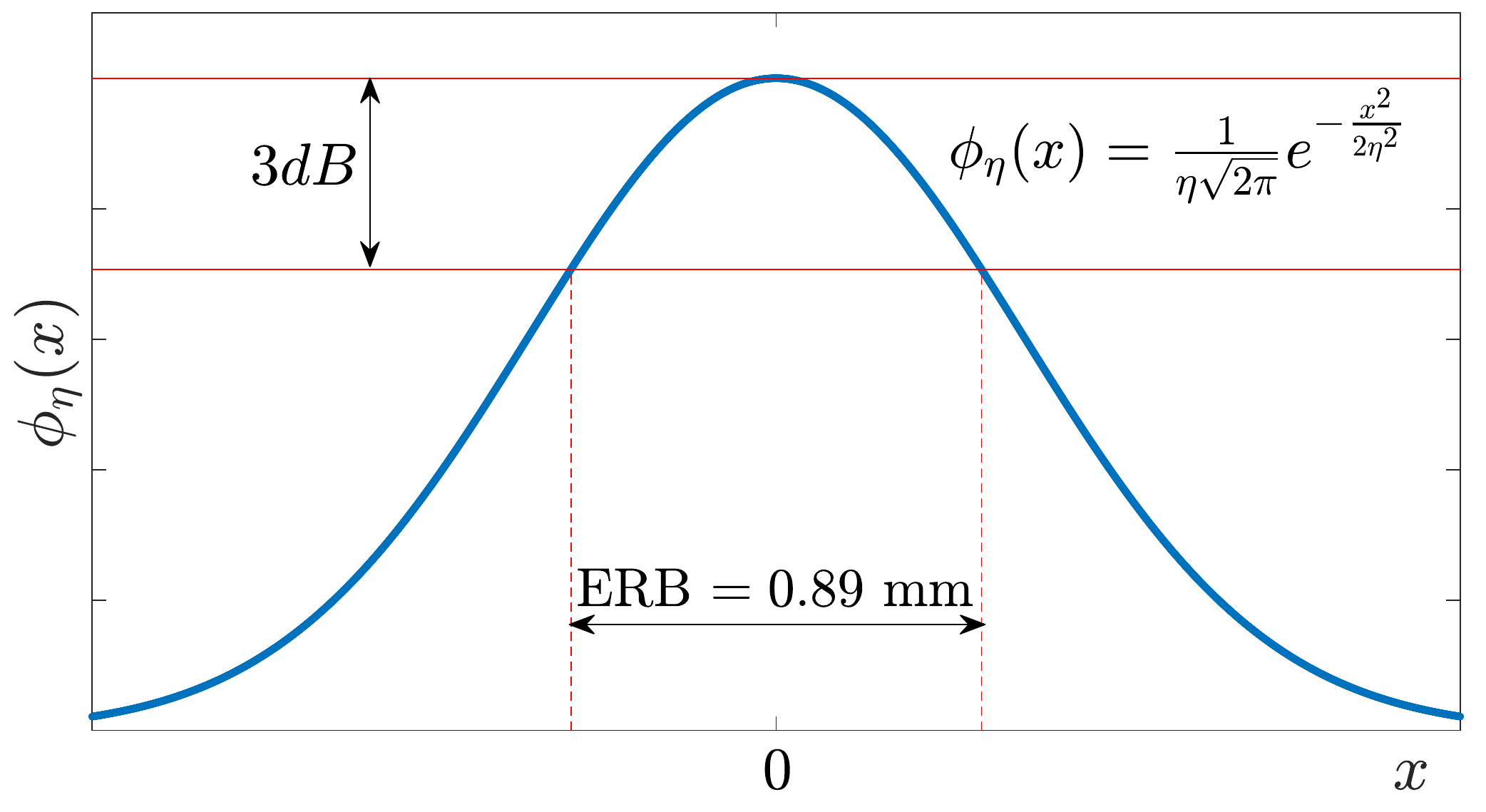} 
	\caption{\footnotesize{Equivalent Rectangular Bandwidth (ERB). The spatial coupling in the micro-mechanical stage is modeled using a Gaussian kernel whose width is chosen to respect the ERB in the cochlea.}}
	\label{Fig: ERB}
\end{figure}

We model the spatial coupling along the BM using a Gaussian kernel as shown in (\ref{Eqn: Nonlinear Gain}-\ref{Eqn: Gaussian Kernel}). Hence, we require to calculate the width $\eta$ of the Gaussian kernel that fits an ERB of $0.89 \si{mm}$ as shown in Figure~\ref{Fig: ERB}. 
It is fairly straight forward to calculate $\eta$, by setting $\phi_{\eta}\left(0.89/2\right)= \frac{\sqrt 2}{2} \phi_{\eta}(0),$
we get $\eta = 0.5345 \si{mm}$.


\end{document}